\documentclass[manuscript]{aastex}
\usepackage{graphicx}

\shorttitle{Heliospheric ENAs from the downwind hemisphere}
\shortauthors{Galli et al.}

\begin{document}

\title{The downwind hemisphere of the heliosphere: Eight years of IBEX-Lo observations}

\author{A. Galli and P. Wurz}
\affil{Physics Institute, University of Bern,
    Bern, 3012, Switzerland}

\author{N. A. Schwadron, H. Kucharek, and E. M\"{o}bius}
\affil{University of New Hampshire, Durham, NH 03824, USA}

\author{M. Bzowski, J. M. Sok\'{o}{\l}, and M. A. Kubiak}
\affil{Space Research Centre, Polish Academy of Sciences,
       Warsaw, 00-716, Poland}

\author{S. A. Fuselier}
\affil{Southwest Research Institute and University of Texas, San Antonio, TX 78228, USA}

\author{H. O. Funsten}
\affil{Los Alamos National Laboratory, Intelligence and Space Research Division, Los Alamos, NM 87545, USA}

\and

\author{D. J. McComas}
\affil{Princeton University, Department of Astrophysical Sciences and Office of the Vice President for the Princeton
Plasma Physics Laboratory, Princeton, NJ 08540, United States}

\begin{abstract}
We present a comprehensive study of energetic neutral atoms (ENAs) of 10 eV to 2.5 keV 
from the downwind hemisphere of the heliosphere. These ENAs are believed to originate 
mostly from pickup protons and solar wind protons in the inner heliosheath. This study 
includes all low-energy observations made with the Interstellar Boundary Explorer over the first 8 years. 
Since the protons around 0.1 keV dominate the plasma pressure in the inner heliosheath in downwind 
direction, these ENA observations offer the unique opportunity to constrain the plasma properties 
and dimensions of the heliosheath where no in-situ observations are available.

We first derive energy spectra of ENA intensities averaged over time for 49 macropixels 
covering the entire downwind hemisphere. The results confirm previous studies regarding 
integral intensities and the roll-over around 0.1 keV energy. With the expanded dataset we 
now find that ENA intensities at 0.2 and 0.1 keV seem to anti-correlate with solar activity. 
We then derive the product of total plasma pressure and emission thickness of protons in the 
heliosheath to estimate lower limits on the thickness of the inner heliosheath. The temporally 
averaged ENA intensities support a rather spherical shape of the termination  shock and a heliosheath 
thickness between 150 and 210 au for most regions of the downwind hemisphere. Around the nominal 
downwind direction of $76^{\circ}$ ecliptic longitude, the heliosheath  is at least 280 au thick. 
There, the neutral hydrogen density seems to be depleted  compared to upwind directions by roughly a factor of 2.
\end{abstract}


\section{Introduction}

The Interstellar Boundary Explorer (IBEX) has been continually observing the interaction
of the heliosphere with the surrounding interstellar medium since 2009 January \citep{mcc09}. The scientific payload
consists of two energetic neutral atom (ENA) imagers, IBEX-Lo \citep{fus09} and IBEX-Hi \citep{fun09}. 
This study expands previous studies of energy spectra of ENAs
observed at low energies \citep{fus14,gal14,gal16}
in terms of time (2009--2016) and spatial coverage 
(the entire downwind hemisphere is included in the analysis and interpretation). The ENAs analyzed here are believed to
originate predominantly from the shocked solar wind and pickup protons beyond the termination shock.
The ENAs thus allow us to sample the source plasma populations over a vast region of the heliosphere 
not accessible to in-situ measurements. 

The ultimate goal of this study is to better understand
the heliosphere, its boundaries, and the properties of plasma populations in the heliosheath.
\citet{mcc13,zir16} investigated the Port Tail and Starboard Tail lobes in IBEX-Hi maps of 
ENA intensities: at solar wind energies or higher, regions of depleted ENA emissions
appear in two lobes $30^{\circ}-100^{\circ}$ apart from the nominal downwind direction. 
Will a similar structure appear at lower ENA energies?
Observations of neutral hydrogen column densities via Lyman-$\alpha$ absorption \citep{woo14}
indicate that the heliotail cannot be deflected by more than 20$^{\circ}$ 
with respect to the nominal downwind direction 
($\lambda_{ecl}=76^{\circ}$, $\beta_{ecl}=-5^{\circ}$) defined by the interstellar flow \citep{mcc15}. 
We will demonstrate in this paper that the low energy ENAs around 0.2 keV sampled with IBEX-Lo 
must be taken into account to understand the complex geometry of the heliosheath in the downwind direction. 
The parent ions of this energy range dominate the plasma pressure. 
With the energy spectra of ENAs measured over the full energy range of IBEX, 
we will be able to derive observational constraints on the geometry, 
cooling lengths, and neutral density of the heliosheath, albeit at a very coarse spatial resolution.  

We present the data set and how we corrected
ENA spectra with the corresponding uncertainties from IBEX-Lo observations (Sections \ref{sec:data} and \ref{sec:errorbars}).  
We then summarize (in Section \ref{sec:results}) the results of the spectra, discuss
the implications of our results on the downwind hemisphere of the heliosphere (Section \ref{sec:discussions}) and 
conclude the paper with a summary and outlook (Section \ref{sec:conclusions}).

\section{Dataset}\label{sec:data}

IBEX-Lo is a single pixel ENA camera \citep{fus09}. Neutral atoms enter the instrument through a collimator, which defines
the nearly conical field-of-view of $6.5^{\circ}$ full width half maximum. A fraction of the incident ENAs then scatter off a
charge conversion surface as negative ions. These ions pass through
an electrostatic energy analyzer and are accelerated into a time-of-flight mass spectrometer, which features
a triple coincidence detection scheme. Apart from negatively ionized ENAs, IBEX-Lo can also detect H$^{-}$ and O$^{-}$
sputtered off the conversion surface by interstellar neutrals (ISN) and ENAs of solar wind energy or higher \citep{par16}.
IBEX-Lo measures ENAs at 8 different
energy passbands with central energies at 0.015, 0.029, 0.055, 0.11, 0.209, 0.439, 0.872, and 1.821 keV \citep{fus09}.

The observation times for this study include the first 8 years of IBEX-Lo triple coincidence 
data of hydrogen ENAs, corresponding to 16 seasonal maps from 2009 January until 2016 October.
Ram maps (in which IBEX, in its orbit around Earth, moves toward the emission source) 
of the downwind hemisphere are created from measurements acquired over April--October each year, 
while anti-ram downwind maps are created from measurements during October--April.
The ram and anti-ram distinction is important to this study because the proper motion of the spacecraft is
not negligible compared to the ENA velocity in the inertial frame. Ram observations therefore have a significantly 
better signal-to-background ratio than anti-ram observations, as discussed in Section \ref{sec:errorbars}.
The dataset includes only the times with the lowest background levels; measurements
affected by high electron background were excluded with the method described earlier by \citet{gal16}.

Two basic limitations of the dataset must be kept in mind. First, IBEX spent the four months
from July until mid-October every year inside the bow shock of Earth's magnetosphere. 
As a consequence, no observations with sufficiently low background level were
obtained. This causes a data gap over
ecliptic longitudes from 0$^{\circ}$ to 120$^{\circ}$, covering most of the downwind hemisphere 
in the ram direction. Ram observations of slow ENAs
have a better signal-to-background ratio than anti-ram observations, which results in smaller error bars 
(see Section \ref{sec:errorbars}). Second, observations after 
2012 July exhibited a lower signal-to-noise ratio: 
the post-acceleration of IBEX-Lo had to be reduced in 2012 July and the background caused by
the terrestrial magnetosphere and the solar wind was elevated from 2012--2016 during the 
maximum of solar activity \citep{gal15}.

We assumed that the energy of detected hydrogen ENAs at 1 au heliocentric distance is the same as their energy
at their place of origin in the heliosheath. 
This is substantiated during solar minimum (2009-2011) because their energy loss, while travelling from the 
plasma source to IBEX, due to solar radiation pressure is nearly compensated 
by the energy gain due to solar gravity, even for low energies at tens of eV \citep{bzo08}.
During solar maximum conditions (2012--2015), a 
22 eV hydrogen ENA emitted far away from the Sun
reaches 1 au at an apparent energy of 15 eV (center of the lowest energy bin of IBEX-Lo), and
 a 36 eV ENA is decelerated to 29 eV (corresponding to the center of energy bin 2) \citep{bzo08}.
For higher energies, the differences are even smaller. We neglected these energy shifts in the study
because we detected no consistent increase of ENA intensity at 15 and 29 eV when we compared ENA maps
from the first 4 years with the maps from 2013--2016.
We also verified for 2009 (solar minimum conditions) and 2014 (solar maximum conditions) that no significant bias 
in corrected ENA intensities occurred at a spatial resolution of $24^{\circ}\times 24^{\circ}$ 
if we replaced our default correction and mapping algorithm by a different algorithm 
based on simulation runs with the Warsaw test particle model \citep{sok15}. In the latter approach, we corrected 
the ENA measurements for the proper motion of IBEX, solar gravity, and the solar radiation pressure.

\begin{figure}
  \includegraphics[trim={0cm 7.3cm 0cm 0cm},clip,width=1.0\textwidth]{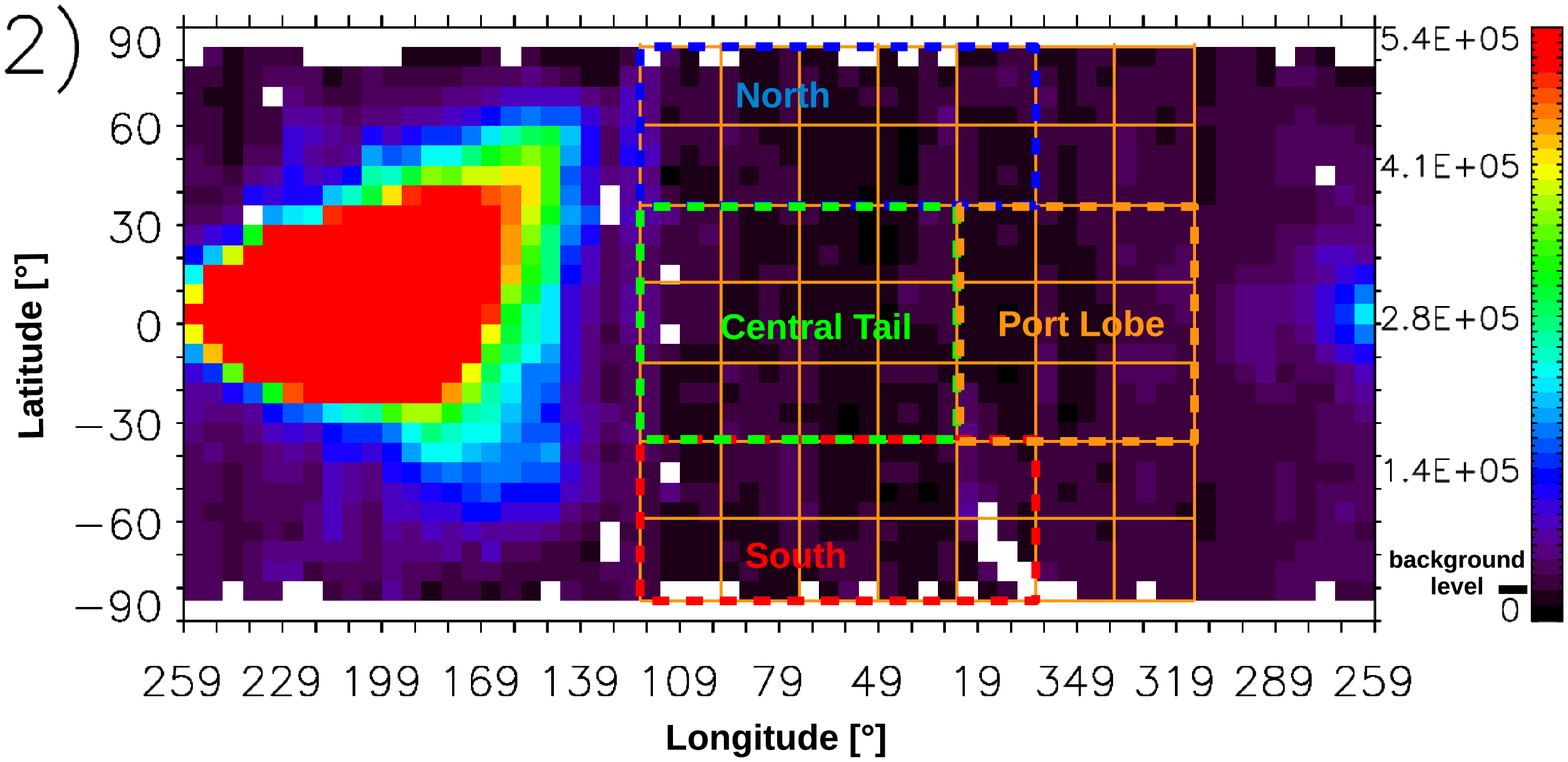}
  \includegraphics[trim={0cm 6.4cm 0cm 0cm},clip,width=1.0\textwidth]{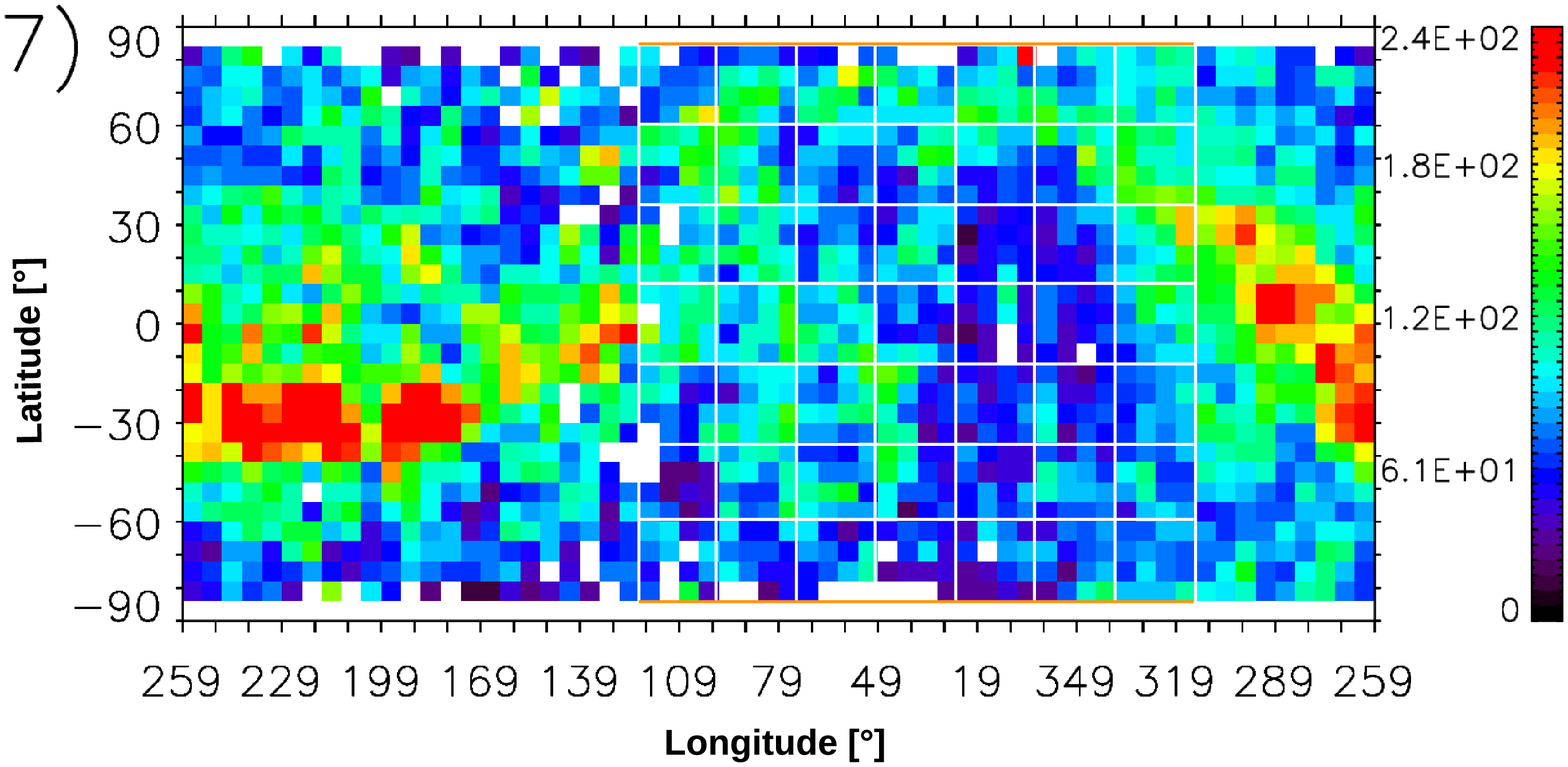}
\caption{High-resolution maps of uncorrected ENA intensities in units of cm$^{-2}$ sr$^{-1}$ s$^{-1}$ keV$^{-1}$ 
in energy bin 2 (top panel, 0.029 keV central energy) and energy bin 7 (bottom panel, 0.872 keV central energy), averaged
from 2009 January to 2016 October. The maps are centered on 79$^{\circ}$ ecliptic longitude. 
The mesh indicates the macropixels covering the entire
downwind hemisphere. Whereas at 0.9 keV most of the ENAs are due to heliospheric ENAs, the intense signal 
at lower energy from upwind directions is due to interstellar neutral helium and hydrogen.
The upper panel also shows the four large regions North (blue), South (red), Central Tail (green), and Port Lobe (orange)
needed for later interpretation of results.}\label{fig:map}
\end{figure}

As in \citet{gal16}, we used data from one single season and one
energy bin and constructed the map of differential
intensities of heliospheric hydrogen ENAs (in units of cm$^{-2}$ sr$^{-1}$ s$^{-1}$ keV$^{-1}$) at 100 au heliocentric distance in the 
inertial reference frame with respect to the Sun at spatial resolution of $6^{\circ}\times6^{\circ}$.
The ENA intensities were first corrected for average sputtering contributions to the ENA measurements and the ubiquitous
background (values as stated in \citet{gal15} for 2009--2012, values for 2013--2016 as stated 
in Table \ref{tab:background_postPACchange}).
After this subtraction, the remaining ENA intensities were corrected for the energy-dependent survival 
probability of ENAs (see Appendices
in \citet{gal16,mcc17}) and for the proper motion of the spacecraft relative to the Sun as described by \citet{gal16}.
We also compensated for the cylindrical distortion in our map projection
for viewing directions at ecliptic latitudes beyond $\pm60^{\circ}$ (see Fig.~\ref{fig:map}).

Using the corrected ENA intensity maps, we then calculated 
the median ENA intensity inside macropixels of size $24^{\circ}\times24^{\circ}$ pixels that are constructed from arrays of 
four $6^{\circ}$ pixels in latitude and four $6^{\circ}$ pixels in longitude.
This mesh of macropixels is shown in Fig.~\ref{fig:map} overlayed on the uncorrected ENA maps
measured at 0.029 keV (top panel) and 0.872 keV (bottom panel). 
At solar wind energy, the ENA intensity map is dominated by the ENA Ribbon \citep{mcc14,sch14}, at low energies, 
it is dominated by the primary and secondary populations of ISN helium and hydrogen \citep{moe12,sau13,kub14,mcc15}.
The macropixels cover the entire downwind hemisphere,
except the polar pixels (studied by \citet{rei16}), with the edges situated at ecliptic longitudes of 120$^{\circ}$, 96$^{\circ}$, 
72$^{\circ}$, 48$^{\circ}$, 24$^{\circ}$, 0$^{\circ}$, 336$^{\circ}$, and 312$^{\circ}$ 
and over the ecliptic latitude range $-84^{\circ}$ to $+84^{\circ}$. The construction of the macropixels was determined by 
balancing the coverage of as much of the downwind hemisphere as possible with equal sized regions, retaining
sufficient spatial resolution to identify variability across and between large emission structures,
and achieving a sufficient signal-to-noise ratio per region for statistically significant results. 
The last criterion meant that we wanted at least $N=10$ ENA counts per season per macropixel at each energy bin
(refer to Section \ref{sec:errorbars} for an explanation).
If less than four of the sixteen $6^{\circ}\times 6^{\circ}$ pixels had a valid intensity, 
for instance because of data acquisition gaps, the macropixel for that season
and energy was omitted from analysis. Zero intensity, on the other hand, was accepted; such cases
occurred at low energies where the measured count rates did not exceed the average background count rates. 
These pixels are colored black in the top panel of Fig.~\ref{fig:map}.
Zero-count pixels provide a threshold sensitivity of the instrument; these results should not be interpreted as 
the absence of ENA emission. We account for this in our error analysis that is described later.

Unless we study the evolution with time, the ENA intensity per macropixel is the median over all available 
seasonal median values, implying 8 ram and anti-ram observations during 2009--2016 for ecliptic longitudes
$0^{\circ}$ to $312^{\circ}$ and 8 anti-ram observations during 2009--2016 for $120^{\circ}$ to $0^{\circ}$.
A median ENA intensity per macropixel must be based on three independent seasonal values,
otherwise that macropixel is omitted from analysis. For interpretation of the results, we grouped the macropixels into 
the four larger regions shown in the upper panel of Fig.~\ref{fig:map}. The value attributed to such a region
is the median intensity calculated over all macropixels.

Contrary to previous studies \citep{gal14,gal16}, we did not a priori exclude any pixels with anomalously high count rate.
Such a cut-off would exclude most of the intense ISN inflow of helium and hydrogen
and would also eliminate some point-like background sources. However, any cut-off criterion for this study 
would be arbitrary and the ENA results might be biased to low intensities
as well as to optimistic uncertainties since the latter were 
computed from the observed variability of intensity. 
 We thus accepted all pixels and verified that
the new median ENA spectrum is identical within error bars to the downwind spectrum in \citet{gal16}
for the same observation period of 2009--2012.
A separation of the globally distributed heliospheric ENA emission and the ISN atoms 
is currently not feasible
because we do not have models of the extended populations of ISN helium and hydrogen
in all IBEX-Lo energy bins. The Warm Breeze model by \citet{kub16}, e.g.,
assumes a simplistic physical scenario. A more sophisticated model of the Warm Breeze, 
originating in the outer heliosheath due 
to charge exchange collisions between interstellar He and He$^{+}$ \citep{bzo17},
is only available for energy bin 2 (around 29 eV).
Fortunately, the ISN inflow hardly affects the results, as will be shown in the results section. 
This non-interference is enabled by our choice of observation direction, which
is restricted to hydrogen ENAs from the 
downwind hemisphere, i.e., looking away from the ISN inflow direction (see Fig.~\ref{fig:map}).

We also investigated signal-to-noise filters as employed by \citet{par16} to exclude
single pixels from uncorrected seasonal maps. However, at low energy most individual pixels have a low
signal-to-noise ratio and the absolute counts per pixel are on the order of 1. As a result, a signal-to-noise
ratio filter was either ineffective or resulted in many excluded macropixels in the
corrected maps. Finally, we investigated the comparison of raw counts in individual macropixels
across different seasons to identify and possibly exclude anomalous pixels before ENA intensity 
correction. However, this was found to be inappropriate because the survival probability at low energies may change by more than 30\%
within one year. We therefore must correct the ENA intensities for survival probability
and the spacecraft motion before we can meaningfully construct and compare macropixel 
averages, variability, and outliers.

\section{Uncertainties and error bars}\label{sec:errorbars}

The corrected ENA intensity of IBEX-Lo measurements is affected both by statistical uncertainty due 
to the small number of counts collected during one season and by systematic errors introduced
by background sources (counts caused by signals other than heliospheric hydrogen ENAs) and by calibration
errors. The first group of uncertainties can be reduced if we average over a larger region
and/or over more seasons of observations. We quantified the various uncertainties associated with ENA intensity
per macropixel the following way:

\begin{enumerate}

\item The pixel-by-pixel standard deviation inside a single macropixel over one season could be due to
a real small-scale spatial variability or due to low count statistics. The latter is more plausible,
as the observed standard deviation agrees with the relative uncertainty expected from a Poisson distribution
with $1/\sqrt{N}$, whereby $N$ equals the number of counts per pixels. This uncertainty increases
markedly for lower energy where fewer counts are available.
The ratios of the pixel-by-pixel standard deviation versus median 
ENA intensity (for the case of an ENA signal distinguishable against background for ram observations) 
calculate to 1, 1.2, 0.8, 0.5, 0.3, and 0.3 for the energy bins at 0.55, 0.11, 0.209, 0.439, 0.872, and 1.821 keV.
The most limiting case at 0.11 keV with an average 0.7 counts (!) per entire season per 
$6^{\circ}\times 6^{\circ}$ pixel explains
why we organized the measurements into macropixels containing 16 single pixels. This way, we achieved
a statistical uncertainty $1/\sqrt{N}$ $<$ 30\% at all energies for a discernible ENA signal.

\item The single count limit must be considered for averaged measurements 
that include multiple zero-count pixels for which both error propagation 
and empirical variability no longer apply: Introducing an artificial map with background count 
rate + 1 count per season ($\approx 0.0015$ cnts s$^{-1}$) everywhere, we assessed the single count limit 
in corrected ENA intensities after corrections for survival probability and Compton-Getting effect: 
Table \ref{tab:singlecountlimits} lists these lower limits imposed by the single count limit.
In the worst case (lowest energy, anti-ram, ecliptic plane), no ENA intensity below
$10^{6}$ cm$^{-2}$ sr$^{-1}$ s$^{-1}$ keV$^{-1}$ can be distinguished against the background level,
even after combining the counts from 16 single pixels! 
For higher energies and for regions outside the ecliptic plane (weaker Compton-Getting effect), 
the single count limit becomes 
irrelevant compared to other uncertainties.
Generally, the single count limit at low energies for a single season agrees with the pixel-to-pixel standard deviation 
presented in point 1. Both uncertainties originate from low count statistics. 

\item To estimate the error of the ENA intensity within a macropixel over 
all different seasons, we relied on the empirical variability between the medians from season to season. 
As previously stated, a minimum of three different seasons was required to calculate a meaningful spread.
Since the seasonal median values are usually not normally distributed (and with some zero values),
we calculated 16\% and 84\% quantiles of seasonal values instead of the simple standard deviation.  
These quantiles represent our best estimate for the $1\sigma$ lower and upper error bar.
They approach the classic standard deviation when the distribution of seasonal values approaches
a Normal distribution. The different seasons constitute statistically independent measurements. Nevertheless,
we did not divide the error bar estimated from the season-to-season difference by the square root
of seasons because the variability may be due to systematic errors (e.g., background sources or time variations) and not only
due to statistical errors. 
The resulting error bars of the averaged signal thus represent the full encountered variability and are rather conservative. 

The medians of the spatial spreads per macropixel reflect the statistical uncertainty of a single season (see point 1). 
That uncertainty becomes negligible compared to the systematic errors when we average over all available seasons.

\item Finally, the absolute calibration uncertainty is $\pm30$\% for a given energy bin. 
This is irrelevant for analysis and interpretation of maps of a single energy bin. 
However, quantitative comparison of intensities across multiple energies 
requires inclusion of this uncertainty. It will be considered the minimum uncertainty for entries
in an energy spectrum.
\end{enumerate}

\begin{figure}
  \includegraphics[trim={0cm 0cm 0cm 0cm},clip,width=1.0\textwidth]{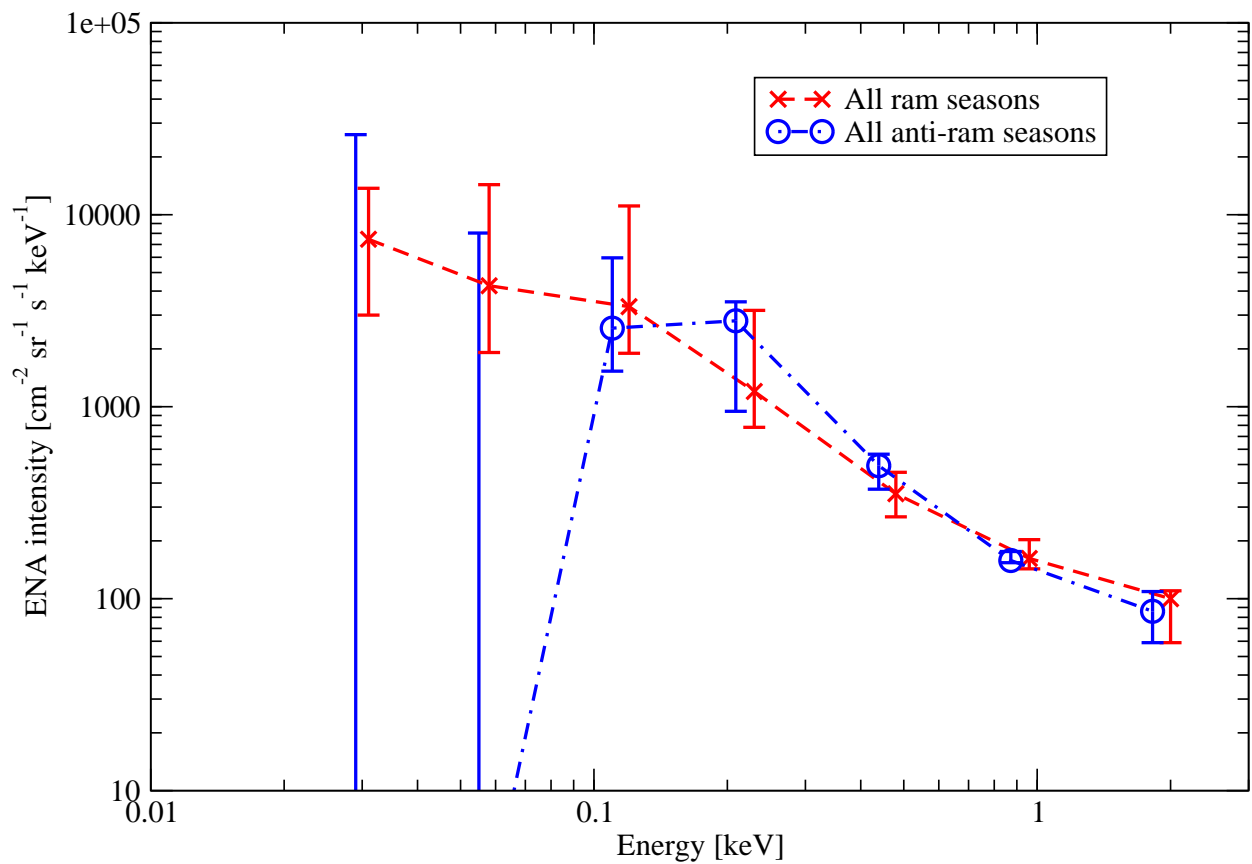}
\caption{Energy spectra of heliospheric ENA intensity in the inertial reference frame at 100 au heliocentric
distance. This figure shows the ENA energy spectrum for the identical macropixel (centered at 
$\lambda_{ecl}=324^{\circ}$, $\beta_{ecl}=48^{\circ}$), 
but sampled in ram (red) and anti-ram (blue) direction.}\label{fig:ram_vs_antiram}
\end{figure}

The default error bars associated with the ENA intensity per macropixel in the following Results
section will be the $\pm 1\sigma$ empirical error bars across different seasons, except if we study temporal evolution
over individual seasons. In that case, the variability per macropixel or per region will serve as an error estimate.
Figure \ref{fig:ram_vs_antiram} illustrates the default error bars.
For this figure we sampled the spectrum at the macropixel centered at 
$\lambda_{ecl}=336^{\circ}\dots312^{\circ}$ and $\beta_{ecl}=36^{\circ}\dots60^{\circ}$, 
which was covered both with ram and anti-ram observations. This example also
demonstrates the challenge of using anti-ram observations at low energies.

\section{Results}\label{sec:results}

Let us first study the spatial distribution and energy spectra of the heliospheric ENAs 
averaged over the entire observation time before
discussing potential temporal trends and contributions by ISN at the lowest energies.

\subsection{Average ENA intensities over all 8 years of measurements}

\begin{figure}
  \includegraphics[clip,width=1.0\textwidth]{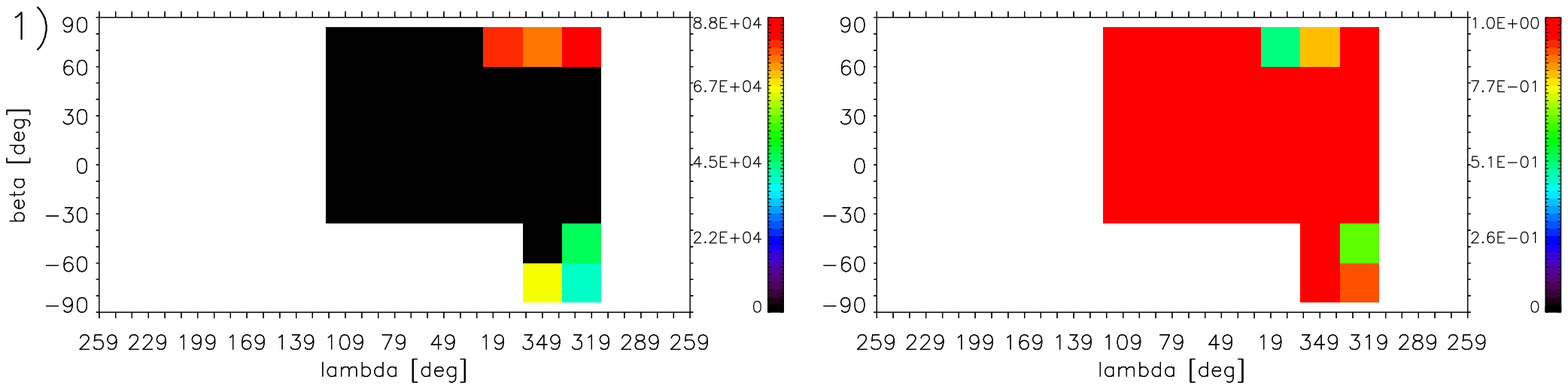}
  \includegraphics[clip,width=1.0\textwidth]{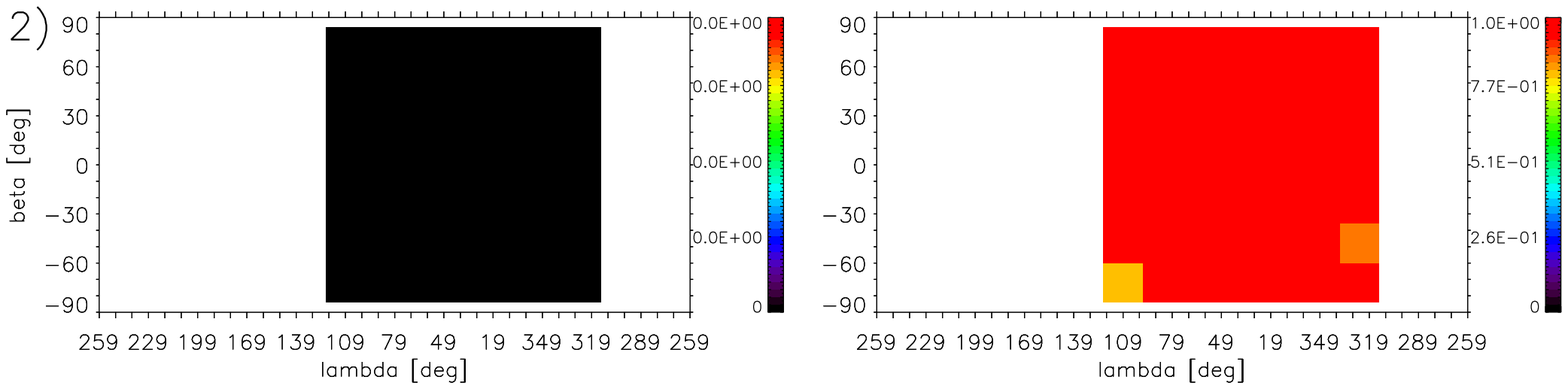}
  \includegraphics[clip,width=1.0\textwidth]{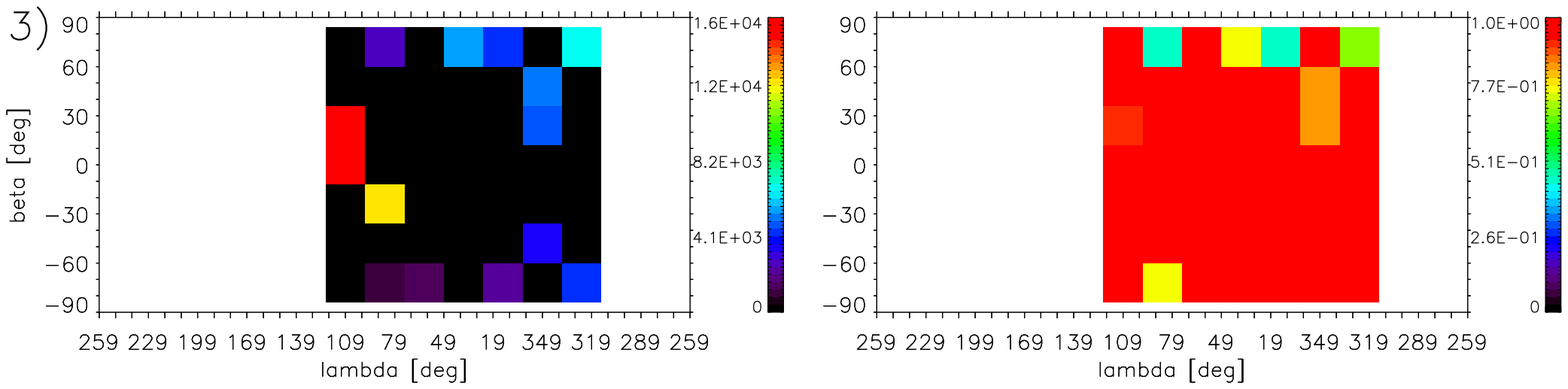}
  \includegraphics[clip,width=1.0\textwidth]{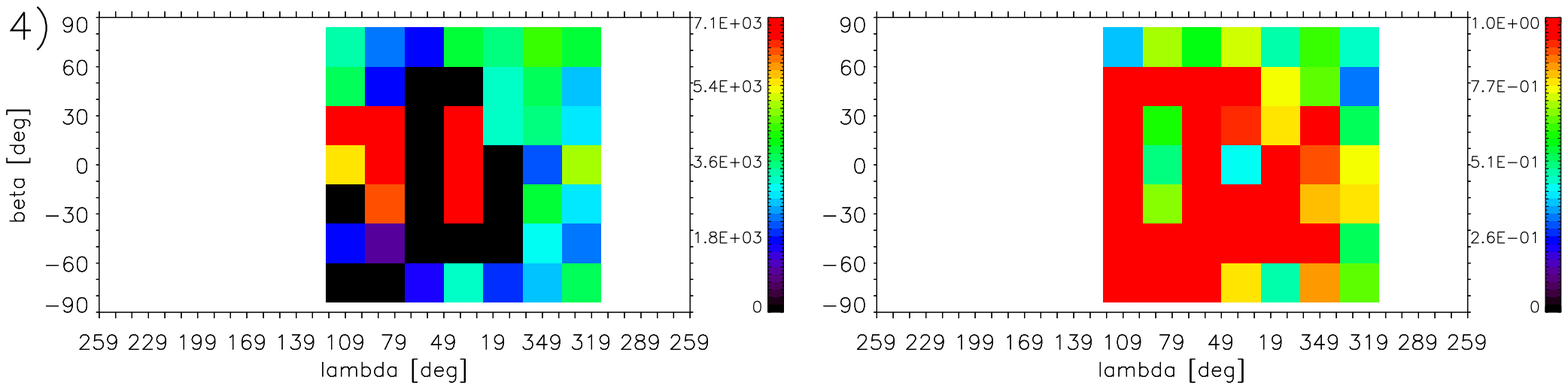}
\caption{Macropixel maps for the lower energy steps 1 to 4 (15 eV to 110 eV) 
for the region of the macropixel mesh defined in Fig.~\ref{fig:map}. Left column:
corrected median ENA intensity over all 16 seasons in units of cm$^{-2}$ sr$^{-1}$ s$^{-1}$ keV$^{-1}$
without background; black pixels indicate median ENA intensities not significantly 
exceeding the background level.
Right column: relative uncertainty.}\label{fig:macropixelmap_1to4}
\end{figure}

\begin{figure}
  \includegraphics[clip,width=1.0\textwidth]{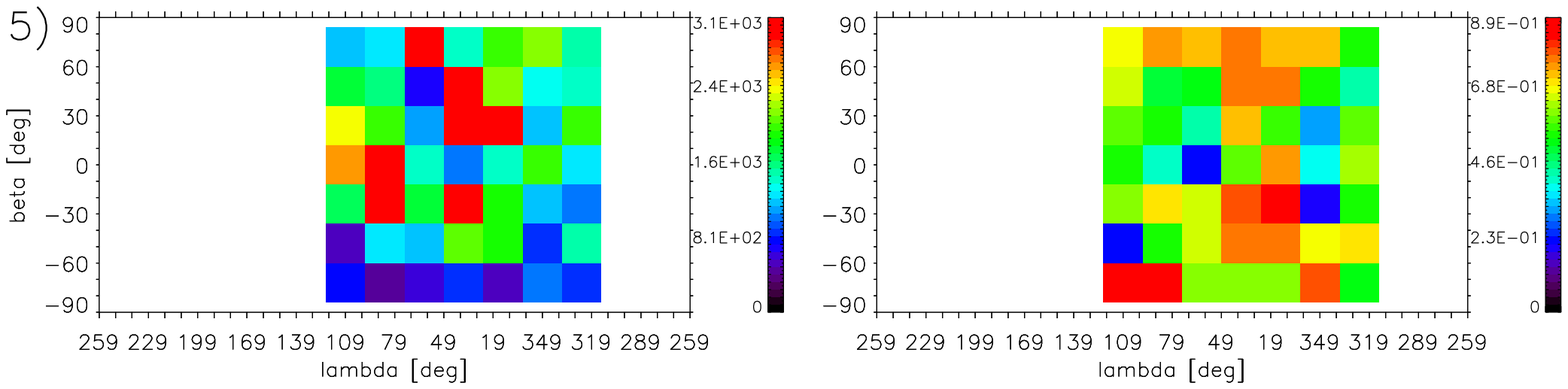}
  \includegraphics[clip,width=1.0\textwidth]{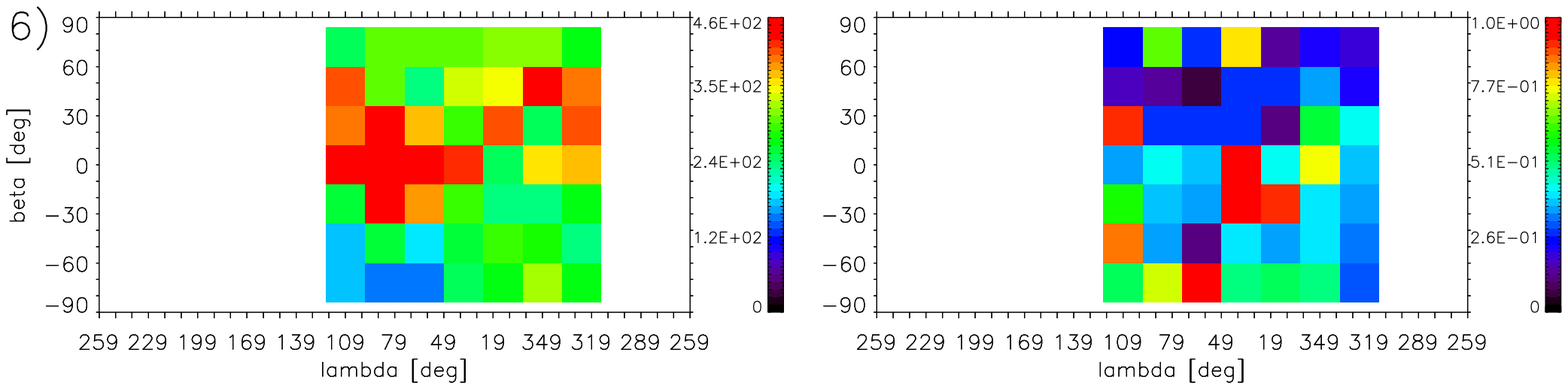}
  \includegraphics[clip,width=1.0\textwidth]{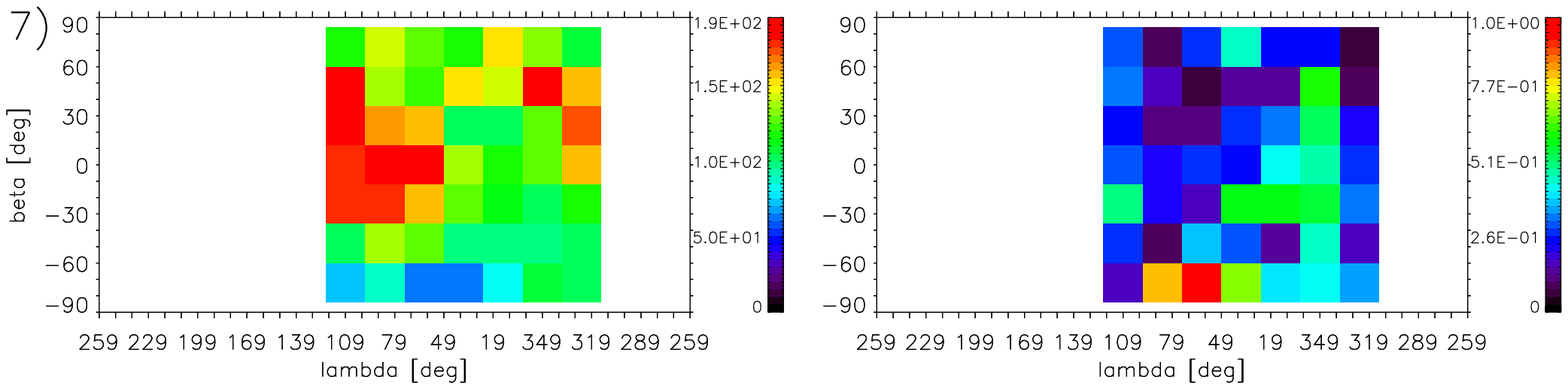}
  \includegraphics[clip,width=1.0\textwidth]{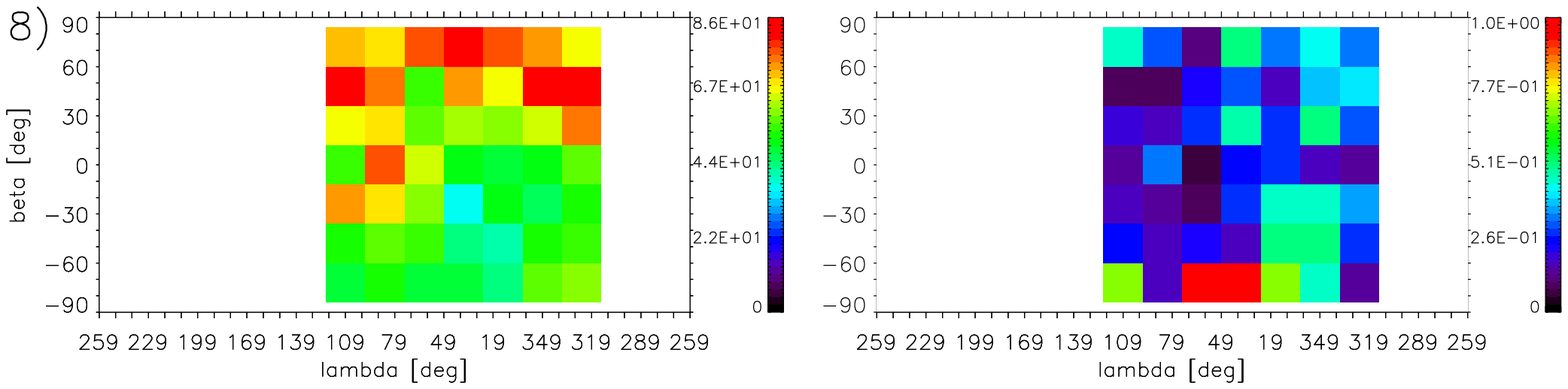}
\caption{Macropixel maps for the higher energy bins 5 to 8 (209 eV to 1.821 keV) 
for the region of the macropixel mesh defined in Fig.~\ref{fig:map}; same format
as Fig.~\ref{fig:macropixelmap_1to4}.}\label{fig:macropixelmap_5to8}
\end{figure}

The left columns of Figures \ref{fig:macropixelmap_1to4} and
\ref{fig:macropixelmap_5to8} show the macropixel maps (defined by the macropixel mesh of Fig.~\ref{fig:map})
of corrected
ENA intensities, averaged over all 16 available seasons. The right columns show
the relative uncertainty ($\sigma_j/j$) of each macropixel; if the intensity is zero, the ratio of 
($\sigma_j/j$) is set to 1. Figure \ref{fig:macropixelmap_1to4} features the 
lower energies (energy bins 1 to 4), while Fig.~\ref{fig:macropixelmap_5to8} shows the higher energies (energy bins 5 to 8).
Even at this coarse resolution, the ENA Ribbon shows up to the North at solar wind energies 
(Fig.~\ref{fig:macropixelmap_5to8}, energy bins 7 and 8).
At intermediate energies (0.1, 0.2, and 0.4 keV) a few pixels of high ENA intensity occur in the ecliptic
but they are interceded with other pixels of low ENA intensity. The only stable spatial feature at 0.1 to 0.9 keV
 is the very low ENA intensity from the South pole region. At energies even lower (bins 1--3 from 0.015 to 0.055 keV),
the downwind hemisphere appears uniformly dim in ENA emission, with few macropixels having
an ENA intensity significantly exceeding
the background level. The exceptions are situated towards the poles because of 
better statistics \citep{rei16} and because
the correction factors due to the Compton-Getting effect and the survival probability are 
notably smaller than at low latitudes.   

\begin{figure}
\begin{center}
  \includegraphics[clip,width=0.75\textwidth]{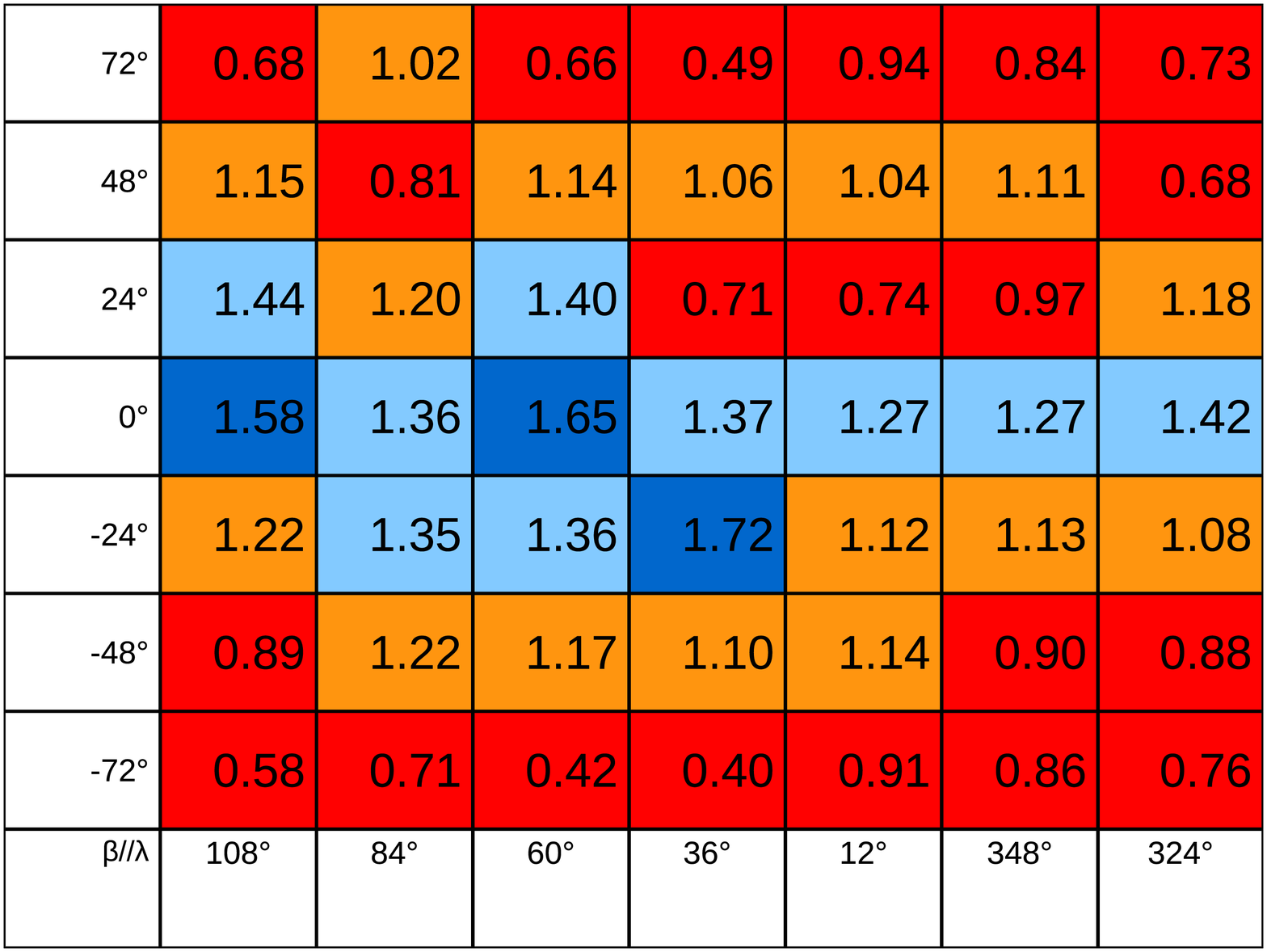}
\caption{Spectral indices $\gamma$ fitted between 0.9 and 1.8 keV
independently for each of the 49 macropixels of this study. Flat energy spectra with small $\gamma$-values are colored
red and orange, steep energy spectra are colored bluish. The ENA energy spectrum is steeper in the ecliptic plane and
flattens towards the poles.}\label{fig:gamma}
\end{center}
\end{figure}

From these maps we calculated the power-law exponent or 
spectral index $\gamma$ of the ENA intensity spectrum above the roll-over energy for each macropixel:
\begin{equation}
j(E) = j_0 (E/E_0)^{-\gamma}
\label{eq:gamma}
\end{equation}
The resulting map of spectral indices between energy bins 7 and 8 (0.9 to 1.8 keV, 
corresponding to solar wind energy) is shown in Fig.~\ref{fig:gamma}.
The spectral index is ordered according to ecliptic latitude
for these energies (in our case $\gamma = 0.3+\cos^{0.8}(\beta)$), whereas
at energies below 0.9 keV the spectral index does not notably vary with latitude.
At the roll-over energy (typically 0.1 keV, see below), the sign of the spectral index changes
and the ENA intensity starts decreasing with decreasing energy. 

The dependence of spectral index on latitude for solar wind energy confirms IBEX-Hi observations by \citet{des15} who found
a similar latitudinal ordering both for upwind and for downwind regions at high ENA energies.
In contrast to the data presented here, \citet{des15,zir17} observed
this latitudinal ordering only above 2 keV. The latitudinal ordering of the energy spectrum
 probably reflects the production mechanism of the corresponding ENAs: ENAs with 1 keV or above
 likely are the neutralized solar wind. Lower energy ENAs originate from other sources in the heliosheath
that have lost the latitudinal ordering of the solar wind speed.
For ENA energies below 0.1 keV, the trace-back times of the parent ions (see following subsection) 
are so long that ENAs measured at IBEX contain a mixture of ENAs generated during
solar maximum and solar minimum conditions. 

At these lower energies, the spectral index of ENA intensities, averaged over all macropixels,
is $1.3\pm0.2$ between 0.4 and 0.9 keV, 
$2.1\pm0.5$ between 0.2 and 0.4 keV, $1.4\pm0.9$ between 0.1 and 0.2 keV, 
and $0.2\pm0.7$ between 0.055 and 0.11 keV, with no notable dependence on latitude or longitude. 
This generalizes the statement in our previous paper \citep{gal16}
that the energy spectrum of ENAs from a few specific downwind regions turns over around 100 eV 
in the whole downwind hemisphere. Unfortunately, we cannot directly compare this roll-over to upwind directions.
There, the inflow of ISN hides the much weaker signal of heliospheric ENAs below 130 eV. 
For the directions towards Voyager 1 and Voyager 2 (upwind hemisphere outside ecliptic plane), 
the observed ENA spectra
appear to be higher than for the downwind hemisphere at low energies (see Fig. 5 in \citet{gal16}). This
implies additional sources of heliospheric ENAs at energies below 100 eV from the
upwind hemisphere, with $j \approx 10^{4}$ cm$^{-2}$ sr$^{-1}$ s$^{-1}$ keV$^{-1}$ at 29 and 15 eV \citep{gal16}. 
However, the lower limit of these ENA intensities is zero
 and the emission intensity is below the instrument threshold, so
we cannot rule out that a similar roll-over
around 100 eV also applies to heliospheric ENAs from the upwind hemisphere.

From here onwards, let us reduce the amount of data to be interpreted. Discussing the spectral shape and temporal trends
of all 49 macropixels independently is not only impractical but also pointless: The uncertainty of the spectral index for a single
macropixel (caused by the uncertainties of ENA intensity at the neighboring energies) is
similar to the fitted value itself even for solar wind energies. Guided by Fig.~\ref{fig:gamma} we thus organized the downwind
 macropixels into just four large regions (see upper panel of Fig.~\ref{fig:map}):
``North'' ($\lambda=120^{\circ} \dots 0^{\circ},\, \beta=36^{\circ} \dots 84^{\circ}$, 10 macropixels), 
``South'' ($\lambda=120^{\circ} \dots 0^{\circ},\, \beta=-84^{\circ} \dots -36^{\circ}$, 10 macropixels), 
and the two regions in the ecliptic plane:
``Central Tail'' ($\lambda=120^{\circ} \dots 24^{\circ},\, \beta=-36^{\circ} \dots +36^{\circ}$, 12 macropixels)
and ``Port Lobe'' ($\lambda=24^{\circ} \dots 312^{\circ},\, \beta=-36^{\circ} \dots +36^{\circ}$, 9 macropixels).
 We separate the Central Tail from the Port Lobe region
because the latter may also contain ISN hydrogen and helium in the two lowest energy bins. 
For the same reason, the four macropixels to the north and south of the Port Lobe region were excluded
from the North and South regions.
The nautical term ``Port Tail lobe'' for the area between $50^{\circ}$ and $0^{\circ}$ at the flank of the heliotail
was introduced by \citet{mcc13} to discuss the shape of the heliotail from IBEX-Hi observations. Heliospheric
ENAs from the opposite region, the starboard lobe at $120^{\circ}$--$150^{\circ}$, could not be distinguished at low energies 
against the inflow of ISN (see upper panel in Fig.~\ref{fig:map}). 

We wanted to expand the energy range of the ENA intensity spectrum for the discussion
of plasma pressure in the heliosheath (see Section \ref{sec:discussions}).
We therefore synthesized and added IBEX-Hi spectra from the most recent Data Release 10 \citep{mcc17}, which represents
7-year averages corrected for Compton-Getting effect and survival probability. 
The energy range for which IBEX-Hi data are evaluated spans $\sim0.5$ to 6 keV
with central energies at 0.71, 1.11, 1.74, 2.73, and 4.29 keV \citep{fun09}. Between 0.5 and 2.5 keV, 
the energy ranges of the two IBEX instruments thus overlap. 
We derived median values and $1\sigma$ variability from the IBEX-Hi Data Release 10 over the four sky regions 
the same way as for IBEX-Lo data.

\begin{figure}
  \includegraphics[clip,width=1.0\textwidth]{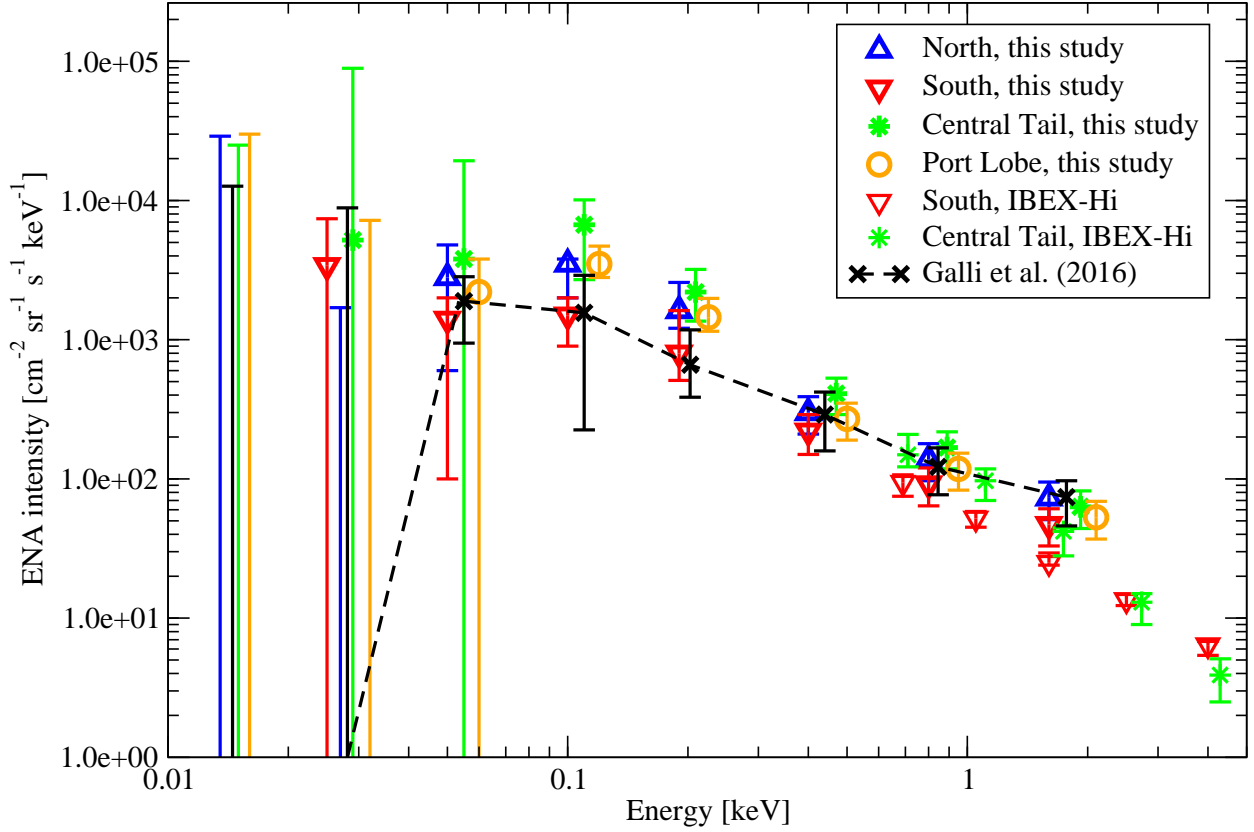}
\caption{Energy spectrum of corrected ENA intensities for four geographic regions in the downwind
hemisphere. ``North'': $\lambda=120^{\circ} \dots 0^{\circ},\, \beta=36^{\circ} \dots 84^{\circ}$, 
``South'': $\lambda=120^{\circ} \dots 0^{\circ},\, \beta=-84^{\circ} \dots -36^{\circ}$, 
``Central Tail'': $\lambda=120^{\circ} \dots 24^{\circ},\, \beta=-36^{\circ} \dots +36^{\circ}$,
``Port Lobe'': $\lambda=24^{\circ} \dots 312^{\circ},\, \beta=-36^{\circ} \dots +36^{\circ}$.
Symbols in bold denote values derived from IBEX-Lo data, thin symbols denote values from IBEX-Hi data.
Black ``x'' symbols show the composite spectrum from a few smaller regions distributed over the
downwind hemisphere \citep{gal16}. For improved clarity, some entries at identical energy were slightly shifted along the x-axis 
and two IBEX-Hi spectra were omitted from the plot.}\label{fig:energyspectra}
\end{figure}

Figure \ref{fig:energyspectra} shows the energy spectra of the four regions; the values are the medians over single
 macropixels. The error bars indicate the 16\% and 84\% quantiles of included macropixel values or 30\% relative error,
whichever is larger. The four energy spectra are compared with the median
IBEX-Hi energy spectra sampled over the same four regions and with the previously published 
``Downwind spectrum'' spectrum over the first 4 years of IBEX-Lo data (black ``x'' symbols, \citep{gal16}).
The colored symbols in bold are IBEX-Lo values, the colored thin symbols are IBEX-Hi values; 
blue triangles up denote North, red triangles down denote South,
green asterisks denote Central Tail, and orange circles denote the Port Lobe.
The spectrum from \citet{gal16} was a composite of four 
smaller areas at the Central Tail and at the northern and southern flanks of the heliosheath 
($\lambda=360^{\circ} \dots 312^{\circ}$), evaluated over the first 4 years of IBEX-Lo observations. 
It matches the 7-years downwind ENA spectra but obviously blurs regional differences.
Another point to note in this figure is that the spectral index of ENA intensities derived with 
IBEX-Hi around 1 keV is steeper than for IBEX-Lo. This bend may be real as ASPERA-3\&4 observations
revealed a knee in the energy spectrum of heliospheric ENAs at $0.83\pm0.12$ keV \citep{gal13}.
Table \ref{tab:energyspectra} lists the IBEX-Lo energy spectra illustrated in Fig.~\ref{fig:energyspectra}.

\subsection{Temporal trends}

\begin{figure}
  \includegraphics[clip,width=1.0\textwidth]{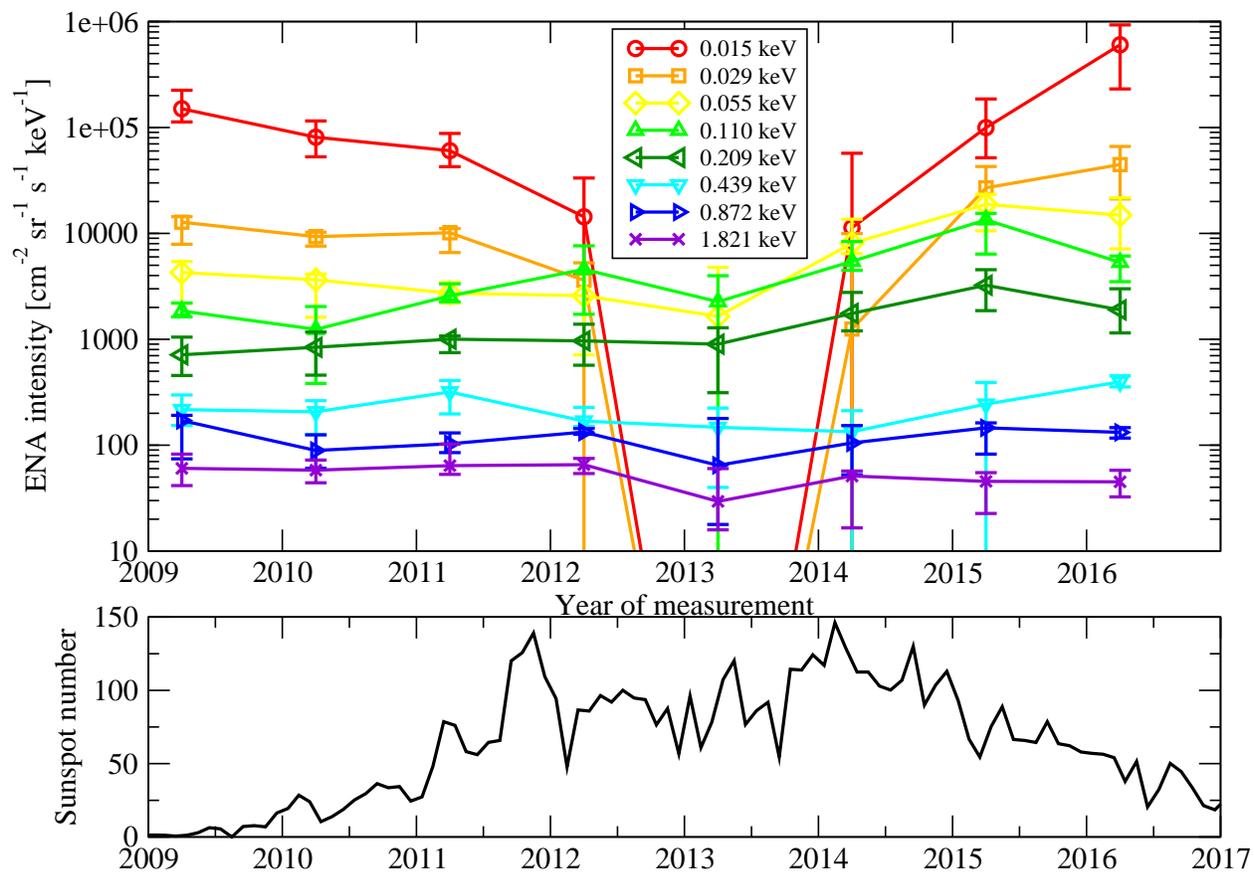}
\caption{Time-series of measured ENA intensities versus solar activity. Top panel: rainbow colored sequence from red (energy bin 1, 0.015 keV) to indigo (energy bin 8, 1.821 keV) 
of ENA intensities measured in the Port Lobe region for individual seasons. Only the 8 ram seasons with their
smaller error bars are shown. Lower panel: monthly averaged sunspot number \citep{wdc08}.}\label{fig:timeseries}
\end{figure}

If we want to compare the ENA signal from individual years
to detect potential trends with time, we can do so only for the Port Lobe
where ram observations were available. The signal-to-noise ratio of anti-ram observations for one single
season proved insufficient (see Section \ref{sec:errorbars}). Moreover, some pixels were not sampled
in every year. 
Figure~\ref{fig:timeseries} shows the timeseries of ENA intensities measured
in the Port Lobe (upper panel) and the solar activity (lower panel). 
Only the 8 seasons of ram observations (April--October) were included. 
Two temporal trends can be discerned in this figure: In the lowest two energy bins, the intensity drops in the years
2011--2013 to background levels before recovering in 2014--2016. This is probably ISN hydrogen
responding to solar activity (see the discussion in Section \ref{sec:isn}). The other notable change
occurs in the years 2014 to 2016 at intermediate energies from 0.05 to 0.2 keV.  
All other temporal variability remains inside the error bars. 

The North, South, and Central Tail regions could only be observed for anti-ram configuration, therefore the 
error bars are generally too large to discern changes between single seasons for energies below 0.1 keV.
At 0.11 keV, on the other hand, the last two of the 8 seasons (October--April 2014/2015 and 2015/2016) 
significantly exceed the median values
also in those three regions. Likewise, the four seasons from 2013--2016
feature significantly higher ENA intensities than for 2009--2012 in all three regions at 0.209 keV.
The apparent dip of ENA intensity around 1 keV in 2013 (purple and blue curves in Fig.~\ref{fig:timeseries}) 
agrees with contemporaneous IBEX-Hi \citep{mcc17} and INCA observations \citep{dia17} 
of the downwind hemisphere. However, the ENA intensities measured with IBEX-Lo at 0.439, 0.872, and 1.821 keV 
in general do not significantly
change over the 8 years in any of the four regions. Figure~\ref{fig:timeseries} is representative 
for the entire downwind hemisphere in this respect. The absence of significant year-to-year variations in
solar wind energy ENAs (0.4 to 2.5 keV) is consistent with the finding of all previous 
studies of IBEX-Lo time-series \citep{fus14,gal14,rei16}. 
IBEX-Hi detected intensity variations with solar cycle on the order of 30\% or less around 1 keV \citep{mcc17}. 
Such variations usually cannot be distinguished with IBEX-Lo because of the poorer signal-to-noise ratio \citep{gal14}.
 
If the observed increase in ENA intensity would occur everywhere between June 2012 and January 2013, 
we might suspect an instrumental effect as the post-acceleration bias in IBEX-Lo changed in that period 
(see Section \ref{sec:data}). However, since the intensity increases only in June 2014 or January 2015 in
several cases and since the intensities at higher energies do not rise at all in later years, a natural
cause seems more plausible. If indeed more ENAs at intermediate energies were produced in the heliosheath
at some time during the solar cycle, we must first derive the trace-back time of the detected ENAs. 
Because temporal variation is likely correlated with the solar cycle, the trace-back time corresponds 
to the time span between the emission of the source plasma (the solar wind) from the Sun, 
its transit to the heliosheath where it becomes ENAs, and the transit time back to the inner heliosphere, 
where the ENAs are detected at IBEX.
Following the notation by \citet{rei12,rei16},
\begin{equation}
t_{tb}(E) = \frac{d_{TS}}{v_{sw}}+\frac{l_{IHS}/2}{v_{ms}}+\frac{d_{TS}+l_{IHS}/2}{v_{\textup{\tiny{ENA}}}(E)},
\label{eq:traceback}
\end{equation}
$t_{tb}(E)$ is the trace-back time for ENAs of
energy $E$ originating in the inner heliosheath, $d_{TS}$ is the distance to the termination shock,
$l_{IHS}$ is the thickness of the inner heliosheath, $v_{sw}$ is the solar wind proton speed, $v_{ms}$
is the average magnetosonic speed in the inner heliosheath, and $v_{\textup{\tiny{ENA}}}(E)$ is
the speed of an ENA of energy $E$ observed at IBEX.
\citet{rei16} estimated $d_{TS}\approx130$ au and $l_{IHS} \approx 210$ au for the North pole,
$d_{TS}\approx110$ au and $l_{IHS} \approx 160$ au for the South pole, and $v_{sw}=690$ km s$^{-1}$ for the polar regions. 
We assigned $l_{IHS}=280$ au and $d_{TS}$ = 110 au to the Central Tail and $l_{IHS}=180$ au and $d_{TS}$ = 120 au to the Port Lobe
(see next Section \ref{sec:discussions}).
The magnetosonic speed in the inner heliosheath \citep{rei12}, 
\begin{equation}
\sqrt{v_s^2+v_A^2}\approx \sqrt{\frac{5}{3}\frac{P}{\rho}}\approx 430 \textup{ km s}^{-1},
\end{equation}
 does not change with latitude if a similar plasma pressure (0.12 pPa = 1.2 pdyn cm$^{-2}$) and proton 
density (640 m$^{-3}$) is assumed for all downwind directions.
Towards Central Tail and Port Lobe, the
solar wind speed inside the termination shock is assumed to be 440 km s$^{-1}$ \citep{wha98}. With these 
speeds and distances, Equation \ref{eq:traceback} yields the trace-back times shown in 
Table~\ref{tab:trace-back} for the 5 cases we need to distinguish. The effects of solar radiation pressure
and gravity on the travel time of low energy ENAs were considered. 
The uncertainties stated in Table~\ref{tab:trace-back} are introduced by the radiation pressure changing with solar activity.

\begin{figure}
\begin{center}
  \includegraphics[clip,width=1.0\textwidth]{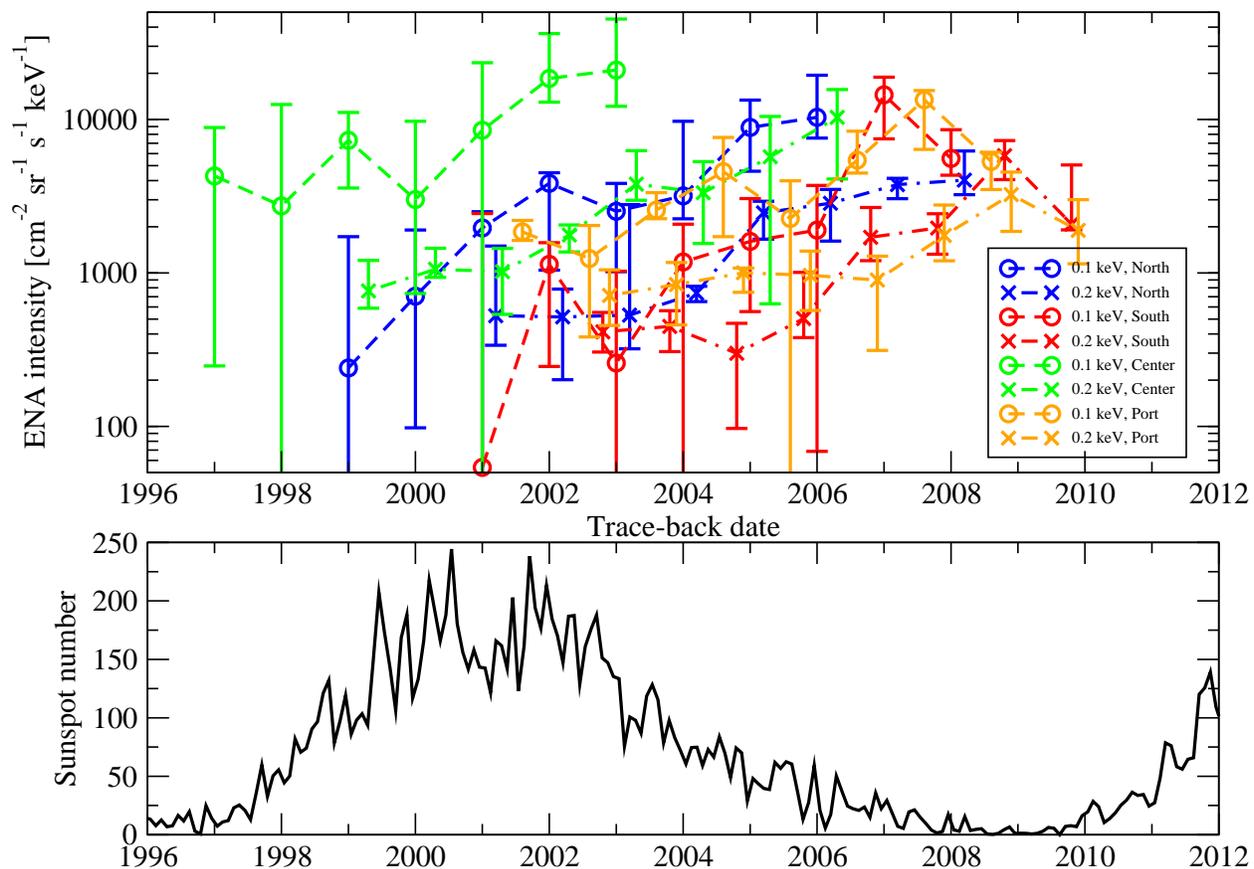}
\caption{Anti-correlation of heliospheric ENAs at 0.1 and 0.2 keV with solar activity. The upper panel
shows the time-series of ENA intensities for all four downwind regions (colored curves)
versus trace-back date, the lower panel shows the monthly mean sunspot number as a proxy for solar activity \citep{wdc08}.}\label{fig:traceback}
\end{center}
\end{figure}

Two things should be noted from this table of trace-back times. First, trace-back dates for ENAs sampled at 15 eV (the lowest energy bin)
from anti-ram direction in the ecliptic plane (Central Tail or Port Lobe) differ by several years depending on whether they were sampled during solar minimum
 (2009) or solar maximum conditions (2014). The incoming ENAs at these low energies are mixed together over an entire solar cycle, 
so the concept of trace-back time characteristic for a given year and region in the sky
cannot distinguish between solar wind emitted during solar maximum and solar maximum. 
Second, time-series of ram and anti-ram measurements at low energies in the ecliptic plane cannot be 
directly compared because their corrected energies, and thus trace-back times, are grossly different. 
This is another reason why we show in Figs.~\ref{fig:timeseries} and \ref{fig:traceback}
only the ram measurements from the Port Lobe region.

The upper panel in Fig.~\ref{fig:traceback} shows the time-series of ENA intensities measured at 0.11 (dashed lines) and 0.209 keV (dashed-dotted lines) 
for all four
downwind regions versus trace-back date. Orange symbols denote Port Lobe (ram observations only), blue symbols
denote North, red symbols denote South, green symbols denote Central Tail (only anti-ram observations available).   
The lower panel of Fig.~\ref{fig:traceback} illustrates the solar activity via the 
monthly mean sunspot number \citep{wdc08}. 
The heliospheric ENA production seems to be anti-correlated with solar activity in
the North, South, and Port lobe region of the downwind hemisphere.
A simple regression analysis reveals that for 7 out of 8 time-series in Fig.~\ref{fig:traceback}
the ENA intensity significantly (confidence levels between 2 to 6 sigma) increases with time from trace-back dates
 1999--2003 (the previous solar maximum) until the end of series at 2006--2010 (solar minimum). The time-series for 0.1 keV at Central Tail
does not follow this trend. 
Towards the North, the ENA increase set in at a trace-back date of 2005;
towards the South and Port Lobe, the ENA increase followed somewhat later in 2006--2007.
 From the anti-correlation of ENA intensity and solar activity, 
we predict that the ENA intensities at 0.2 keV
will remain high until the middle of 2017 before decreasing again when the trace-back date will
correspond to increasing solar activity. The decrease of 0.1 keV ENA intensity should follow one to two years later.
No conclusion can be reached yet with respect to the time variability of the Central Tail. The
 ENA intensities at 0.1 keV (green dashed line) linked to a trace-back date before 1999 show no tendency to increase with low solar activity. 

At 0.4 and 0.9 keV, only 1 out of the 8 time-series from the downwind regions exhibits a slope significantly different from zero. 
In all other cases, a constant fits all 8 seasonal values within respective error bars.
The variations of ENA intensity with time are generally less pronounced at solar wind energies than for 0.08--0.3 keV 
and for 3--6 keV (see \citet{mcc17} about the temporal evolution over the full sky and \citet{rei16} 
regarding the polar regions). With 8 years of observations available, both IBEX-Lo and IBEX-Hi
can be used to track the imprint of the solar cycle on the heliosheath plasma. 





We cannot answer yet which physical process causes the anti-correlation between solar activity and production of 
0.1--0.2 keV ENAs in the heliosheath. But we have seen that the temporal evolution
of ENA intensities looks similar for all directions in the downwind hemisphere with the possible exception of the Central Tail. 
This implies via the trace-back time in Equation \ref{eq:traceback} that the region of ENA production 
has a similar distance towards the poles
and towards the flanks of the heliotail. The estimate for the Central Tail may differ from our assumption. 
In Section \ref{sec:discussions} we will motivate these assumptions by deriving
the heliosheath thickness from the total plasma pressure averaged over time.
Once the IBEX-Lo time-series cover an entire solar cycle, we can use an updated version of Fig.~\ref{fig:traceback}
to optimize the trace-back times and from there the travel distance of ENAs from the downwind hemisphere.

\subsection{Interstellar neutral hydrogen observed at the lowest energy bins}\label{sec:isn}

The ISN inflow was seen to extend to longitudes $>300^{\circ}$ in the ecliptic plane in the overview figure
of energy bin 1 (Fig.~\ref{fig:map}). Looking at the temporal evolution in Fig.~\ref{fig:timeseries} 
for the Port Lobe, we
recognize the ISN signal in energy bin 1 in 2009--2011, then it vanishes in 2012, only to reappear in 2014.
Remember that those ENA measurements include only the ram observations. The anti-ram measurements
of the same sky direction usually yield no detectable ENA signal 
(see the black macropixels at 15 eV and 29 eV in Fig.~\ref{fig:macropixelmap_1to4}). 
Heliospheric ENAs of 15 eV with trace-back times of decades obviously cannot produce a bi-annual intensity change,
so we conclude that this is indeed the outermost part of the seasonal ISN inflow.
 
The energy of ISN hydrogen is too low to produce any signal above energy bin 2, and ISN helium produces a strong signal
in IBEX-Lo in all 4 energy bins below 150 eV \citep{sau13}. This can be easily understood from the energy 
of the ISN particles entering IBEX-Lo. Whereas ISN helium has a maximum relative energy of 130 eV
\citep{gal15} for ram measurements, ISN hydrogen will at most have the same velocity as ISN helium but, having
4 times less mass, only reach 33 eV. This is close to the central energy of bin 2 at 29 eV. Any non-gravitational
influence on the ISN hydrogen trajectory, such as solar radiation pressure, will tend to further slow down
hydrogen with respect to helium. As the signal temporarily disappears only
in energy bins 1 and 2 but not in energy bins 3 and 4 in Fig.~\ref{fig:timeseries}
we conclude that this signal is mostly ISN hydrogen. In the ecliptic plane it
extends to an apparent direction of $336^{\circ}\pm6^{\circ}$, which is even wider than the Warm Breeze 
of the secondary helium \citep{kub16}. The latter blends into background around $300^{\circ}\pm6^{\circ}$, judging
from count rate maps in analogy to Fig.~\ref{fig:map} for 0.055 keV.

The variation of the ISN hydrogen signal from 2009 to 2016 in Fig.~\ref{fig:timeseries} is probably caused
by the varying survival probability for neutral H reaching the inner solar system depending on solar cycle. 
\citet{sau13} found that this hypothesis explained the IBEX-Lo observations of ISN hydrogen during
 the first part of the solar cycle. 
The rapid recovery of the ISN hydrogen in 2014 is puzzling, however. 
Based on the solar activity (see Fig.~\ref{fig:timeseries}) we would have expected the signal to recover
only in 2016. 
Possibly, the model used to estimate hydrogen ENA losses due to re-ionization close to 1 au is not accurate enough 
at these very low energies. 
We cannot verify if the heliospheric ENAs are affected the same way, i.e., whether they also rise rapidly in 2014 
in all other downwind regions at 15 eV and 29 eV. 
At these energies, the anti-ram observations do not allow for a meaningful seasonal analysis.
We will devote a future study to this ISN hydrogen signal and how it expands and diminishes over a full
solar cycle. For the discussion of heliospheric ENAs in the following section we will disregard the
ISN contribution to the Port Lobe below 0.055~keV.   

\section{Implications for the heliosheath in the downwind direction}\label{sec:discussions}

Throughout this discussion we assume that the observed ENAs originated exclusively in the inner heliosheath.
Contributions from the outer heliosheath are unlikely because
of the increasing heliosheath thickness in downwind direction. The ENA intensities observed with IBEX-Lo from 
Voyager 1 and Voyager 2 direction in the upwind hemisphere can be reproduced without a contribution of ENAs from 
the outer heliosheath \citep{gal16}. It is therefore unlikely that ENAs from the outer heliosheath
contribute notably to the ENA intensity in the downwind hemisphere.

The ENA measurements provide insight into the integrated plasma pressure over the 
line-of-sight thickness of the source plasma population from which the ENAs are emitted. 
We repeat the plasma pressure calculation presented by \citet{sch11} (also see \citet{fus12,sch14,rei12,rei16,gal16})
for the new ENA energy spectra averaged over the four regions.
Because this is based on observations, the integration is done step-wise over each
of the energy bins of the instrument; the reference frame of the plasma pressure is heliocentric, i.e.,
not moving with the plasma bulk flow speed $u_R$ \citep{sch11}:

\begin{equation}
\Delta P \times l = \frac{2 \pi m^{2}}{3n_{H}} \frac{\Delta E}{E} \frac{j_{\textup{\tiny{ENA}}}}{\sigma(E)} \frac{(v_{\textup{\tiny{ENA}}} + u_R)^4}{v_{\textup{\tiny{ENA}}}}
\label{eq:pressure1}
\end{equation}
\noindent The measured intensity $j_{\textup{\tiny{ENA}}}$ of neutralized hydrogen at a given energy 
thus translates into the product $\Delta P \times l$ of the pressure of the parent ion population in the heliosheath
that is the ENA emission source and the thickness of this ion population source region along the instrument line-of-sight. 
This pressure includes only the internal 
pressure of the moving plasma, there is no ram pressure term contributing to the balance for the downwind hemisphere. 
Equation \ref{eq:pressure1} states that the product of pressure times ENA emission thickness can be derived
from observations, but to obtain the two values separately further assumptions will be needed.
The equation can be rewritten in the notation of \citet{fus12}
as the product of a stationary pressure (the internal pressure in the inertial reference frame with $u_{R}=0$) 
times a correction factor for the plasma bulk flow velocity with respect 
to the heliocentric rest frame:
\begin{equation}
\Delta P \times l = \frac{4 \pi m_{H}}{3 n_{H}} \frac{v_{\textup{\tiny{ENA}}} j_{\textup{\tiny{ENA}}}(E_0)}{\sigma(E_0)} \int^{E_0+\Delta E/2}_{E_0-\Delta E/2} dE \left(\frac{E}{E_0}\right)^{-\gamma} \, c_f
\label{eq:pressurebalance}
\end{equation}
\begin{equation}
c_f = \frac{(v_{\textup{\tiny{ENA}}} + u_R)^4}{v_{\textup{\tiny{ENA}}}^4}.
\label{eq:correctionfactor}
\end{equation}
In Equation \ref{eq:pressurebalance}, $\Delta E$ denotes the width of the respective energy bin and $\gamma$
is the spectral index (see Equation \ref{eq:gamma}).
For the typical radial velocity of solar wind in the downwind hemisphere of the inner heliosheath, we assumed
$u_R=140$ km s$^{-1}$ everywhere, as measured by Voyager 2 \citep{wha99,sch11} and similar to the range of 100--150 
km s$^{-1}$ assumed by \citet{zir16} over the first 100 au beyond the termination shock.
More precisely, $u_{R}$ would be a function of heliolatitude with faster plasma speeds -- up to 225 km s$^{-1}$ -- 
occurring towards the poles \citep{rei16}. Such a latitudinal dependence could be formulated 
if we assumed a shock jump of 2.5 everywhere \citep{sch11}. But for this discussion let us assume a constant plasma bulk 
flow speed for the entire downwind hemisphere. As in previous IBEX-related papers on pressure in the heliosheath,
let us first assume a constant density of neutral hydrogen everywhere in the inner heliosheath 
with $n_H=0.1$ cm$^{-3}$ \citep{sch11,glo15}. We will re-assess this assumption at the end of this section. 
The charge-exchange cross section $\sigma(E_{0})$ between protons and neutral hydrogen 
is taken from \citet{lin05}, decreasing from 4 to 2 $\times10^{-15}$ cm$^{-2}$ as the ENA energy increases 
from 0.015 to 1.821 keV. We applied Eq.~\ref{eq:pressurebalance} to the average ENA intensities 
in Figure~\ref{fig:energyspectra} and Table~\ref{tab:energyspectra} to calculate stationary and dynamic pressure for all regions.
In the stationary case, the ENAs at solar wind energies would dominate the total pressure balance.
For the following discussions and illustrations, however, we will concentrate on the dynamic pressure because
the plasma is flowing away from the Sun.  

\subsection{Plasma pressure in the heliosheath}

\begin{figure}
  \includegraphics[clip,width=1.0\textwidth]{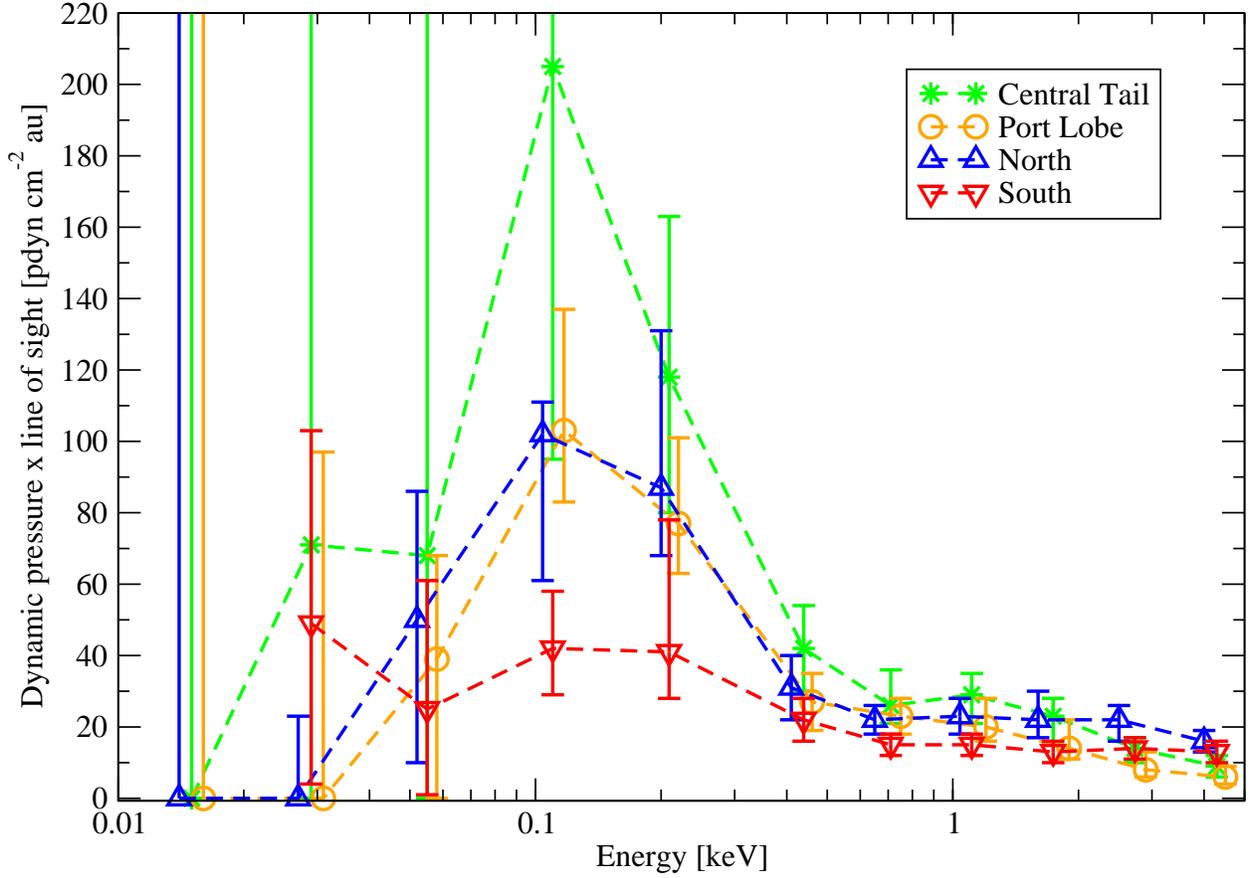}
\caption{Energy spectrum of $\Delta P \times l$ (partial dynamic pressure times ENA emission thickness) in units of pdyn cm$^{-2}$ au. 
The symbols have the same meaning as in Fig.~\ref{fig:energyspectra}, but here IBEX-Lo 
and IBEX-Hi data have been combined into composite spectra from the 6 lower energy bins of IBEX-Lo and the 5
higher energy bins of IBEX-Hi.}\label{fig:pressurespectrum}
\end{figure}

To obtain the full plasma pressure $P$ in the heliosheath we would like to integrate the ENA spectrum from zero to
infinite energy. As will be shown briefly hereafter, IBEX-Lo usually covers most energies at the lower end that
contribute to the total pressure. To assess how much pressure would be added at energies above the IBEX-Lo cut-off at 2.5 keV, 
we also considered the IBEX-Hi spectra averaged over the four regions (see Tab.~\ref{tab:energyspectra}).
As in our previous study \citep{gal16}, the relative uncertainty attributed to IBEX-Hi 
measurements is 20\% or the standard deviation between the four different
regions, whichever is larger.
Whereas the spectral index derived from IBEX-Hi spectra is steeper than for IBEX-Lo spectra at the overlapping 
energy, the dynamic pressure added over the two IBEX-Lo bins 7 and 8
(covering roughly 0.6--2.5 keV), agrees well with the dynamic pressure integrated over IBEX-Hi energy bins 2 to 4 (0.5--2.5 keV):
(72 and 67) pdyn cm$^{-2}$ au for the North, (47 and 40) pdyn cm$^{-2}$ au for the South,
(74 and 78) pdyn cm$^{-2}$ au for the Central Tail, and (57 and 50) pdyn cm$^{-2}$ au for the Port Lobe.
We therefore combined the 6 lower energy bins of IBEX-Lo 
with the upper 5 energy bins of IBEX-Hi to obtain four composite energy spectra spanning the whole range from 0.01 to 6 keV.
These pressure spectra are averages over all 8 years (IBEX-Lo) or 7 years (IBEX-Hi) of observations.
The resulting $\Delta P \times l$ per individual energy bin is shown for all four
regions in Fig.~\ref{fig:pressurespectrum} (orange triangles up: North, green triangles down: South,
red asterisks: Central Tail, blue circles: Port Lobe). The error bars
represent the variability within a region; the uncertainties of the plasma bulk flow velocity were neglected.

The values of the pressure spectra plotted in Fig.~\ref{fig:pressurespectrum} are stated 
in Table \ref{tab:pressurespectrum}. Table \ref{tab:pressure} gives the corresponding 
total $P \times l$ over two energy ranges (0.01 to 6 keV and 0.08 to 6 keV).
The ENA energies at 0.11 keV and 0.209 keV
contribute most to the total pressure for all four downwind regions.
We therefore set 150 eV as the relevant energy to discuss the cooling length \citep{sch11,sch14} 
in the downwind heliosheath.     
The energies below the roll-over around 100 eV contribute on average little to the total pressure 
because most macropixels
feature no discernible ENA signal above the background (see Fig.~\ref{fig:macropixelmap_1to4}). 
However, because of the large upper error bars and the large correction 
factor at low ENA energies ($c_f=170$ at 15 eV), the Central Tail could feature 
a high upper limit of 1800 pdyn cm$^{-2}$ au
if the energy range is extended down to 10 eV. We therefore added the plasma pressure
over two different energy ranges in Table \ref{tab:pressure} to indicate the uncertainties
at the lowest ENA energies.

Before interpreting the derived values of $P \times l$ we would like to caution the reader about the effect of solar
activity on the values in Table \ref{tab:pressure}. That table incorporates all available observations, thus blending
ENA observations representative for high and low solar activity conditions in the heliosheath. 
Because IBEX observations do not yet cover an entire solar cycle, we cannot create separate
plasma pressure spectra for low and high solar activity
over the entire energy range of IBEX-Lo and IBEX-Hi.
The trends of ENA intensities at 0.1 and 0.2 keV in Fig.~\ref{fig:timeseries} 
indicate that the $\Delta P \times l$ decreases by a factor
of 3 during high solar activity with respect to the average stated in Table \ref{tab:pressure}
and increases by a similar factor for low solar activity conditions. Because ENAs of these energies
are a dominant contribution to the total plasma pressure (see Fig.~\ref{fig:pressurespectrum}),
$P \times l$ varies strongly between the two extreme cases: we estimate 
that $P \times l$ increases from $300\pm100$ to $1050\pm300$ pdyn cm$^{-2}$ au in the Central Tail
(integrated from 0.08--6 keV) as the heliosheath changes from high to low solar activity conditions.
The IBEX-Lo observations of 2017 and 2018 will inform whether this anti-correlation
with solar activity is real: these years combine a low present solar activity, 
which allows for a better signal-to-noise ratio, with a trace-back time for 0.1 keV ENAs corresponding 
to the previous solar minimum. However, we can state with confidence the lower limits of $P \times l$
over the solar cycle. We assumed the lower limits in Table \ref{tab:pressure} and reduced the contributions
at 0.1 and 0.2 keV according to the temporal trend in Fig.~\ref{fig:timeseries}. The results are representative
for high solar activity conditions and optimum signal-to-noise ratio during the first 3 years of IBEX-Lo
observations. These lower limits, evaluated from
0.01 to 6 keV, read: 210, 150, 280, and 180 pdyn cm$^{-2}$ au 
for the North, South, Central Tail, and the Port Lobe, respectively.

The main challenge with interpreting the product $P \times l$ is disentangling
the two factors. If they are independent and thus separable, a governing assumption is that $P$ is constant over $l$.
 Both parameters are unknown a priori, as no in-situ measurements are available, 
contrary to the upwind hemisphere observed by the Voyager spacecraft \citep{bur08,dec05,gur13,sto13}. If the plasma pressure is assumed to be similar over 
all directions in the heliosheath, the numbers in Table \ref{tab:pressurespectrum}
translate directly into the thickness of the emission structures along lines of sight in the heliosheath. 
But what should that total pressure be?

Most studies, no matter if they are based on models or observations, predict a total plasma pressure
of 1--2 pdyn cm$^{-2}$ in the inner heliosheath: 
\citet{liv13} derived a total pressure of 2.1 pdyn cm$^{-2}$ over the entire energy spectrum, 
comparing the expected plasma
pressure from a kappa distribution of protons with the plasma pressure derived from IBEX-Hi energy spectra.
These authors found
the same value for all sky directions except for the ENA Ribbon towards the nose of the heliosphere.
Rather at the lower limit, \citet{rei12} and \citet{glo11} derived a total plasma pressure
of 1.2 pdyn cm$^{-2}$ from Voyager and early IBEX observations. 
This pressure is dominated by the heliospheric pickup ions, 
which contribute $1.0 \pm 0.5$ pdyn cm$^{-2}$ \citep{glo11}.
\citet{gal16} derived 1.4 pdyn cm$^{-2}$ for a few regions in the flanks and in the heliotail 
for the IBEX-Lo energy range from 10 eV to 2.5 keV.
\citet{rei16} derived a heliosheath thickness of 160 au for
 the South pole region and 210 au for the North pole region, using a different approach
of analyzing the temporal variability of heliospheric ENAs at several keV. These dimensions imply 
with our values for $P \times l$ in Table~\ref{tab:pressure}: 
$P = 1.8+1.6-0.3$ pdyn cm$^{-2}$ from 0.01 to 6 keV to the North, which equals
the $1.6\pm0.4$ pdyn cm$^{-2}$ to the South. This reduces to $P = 1.3\pm0.2$
pdyn cm$^{-2}$ from 0.08 to 6 keV for the two polar regions.
Applying $P = 1.3$ pdyn cm$^{-2}$ to Table~\ref{tab:pressurespectrum}, we then find
 $l=215+35-22$ au for the ENA emission thickness towards the Port Lobe and $360\pm85$ au towards
the Central Tail. The lower limits of $P \times l=$ 210 and 150 pdyn cm$^{-2}$ au
imply $P = 1.0$ pdyn cm$^{-2}$ and hence the heliosheath must be at least 280 au in the 
Central Tail and 180 au towards the Port Lobe for any time during the solar cycle.

The difference of $P \times l$ observed towards North and South confirms the 
dichotomy found by \citet{rei16} at higher energies (0.5 to 6 keV). 
This independent confirmation at lower energy indicates that 
the downwind heliosheath likely extends farther to the North than to the South.
The dichotomy already appears in the stationary pressure; the lines of sight would be similar at both poles
only if the radial bulk flow velocity towards the South were much higher 
($u_R\approx 200$ km s$^{-1}$ instead of the assumed 140 km s$^{-1}$) than towards the North
to compensate for the observed two-fold difference via the dynamic correction factor (see Eq.~\ref{eq:correctionfactor}).

The minimum ENA emission thickness in the Central Tail region
considerably exceeds the lower limits encountered
for the other downwind regions (high latitudes or the flank of the heliosheath around $\lambda=0^{\circ}$.
The presumed tail of the heliosheath therefore is not visibly deflected from the
nominal downwind direction of $76^{\circ}$ ecliptic longitude \citep{mcc15}.   
This confirms Lyman-$\alpha$ observations by \citet{woo14} of interstellar absorption
towards nearby stars in the downwind direction.
The Port Tail and Starboard Tail lobes show up beside the nominal downwind direction
in ENA maps above 1.5 keV energy, but these lobes then seem to blend into the globally distributed
ENA flux at solar wind energy \citep{mcc13,zir16}. A similar bi-lobate structure at 
lower energies has not been detected so far. 
Since the total plasma pressure is dominated by those lower energies, the IBEX data
are consistent with a symmetric heliotail not notably offset from the nominal downwind direction. 

\subsection{Plasma cooling length and the dimension of the heliosheath}

\citet{sch11} introduced the concept of a cooling length $l_{cool}$ in the inner heliosheath 
 as the time scale for plasma ions being neutralized times the plasma bulk flow velocity:
\begin{equation}
l_{cool} = \frac {u_R}{n_H\sigma v_{\textup{\tiny{ENA}}}}
\label{eq:coolinglength}
\end{equation} 
The mean free path length a hydrogen ENA of the same energy 
can travel before being lost to re-ionization is much longer than that. 
\citet{sch11} derived from Equation~\ref{eq:coolinglength} a cooling length of 120 au in the inner heliosheath 
for plasma sampled with ENAs of 1 keV energy, which is appropriate for the upwind heliosheath. 
For the downwind hemisphere, the energy of 150 eV, relevant for the total pressure 
(Fig.~\ref{fig:pressurespectrum}), implies
 $v_{\textup{\tiny{ENA}}}= \sqrt{2E/m} \approx 170$ km s$^{-1}$ and thus $l_{cool}=210$ au.
For the lowest energy bin of 15 eV, the cooling length would increase to 430 au.

A similar mean free path is found for any heliosheath proton moving through the surrounding
neutral hydrogen. The pick-up proton density and the solar wind density
are both on the order of only $10^{-3}$ cm$^{-3}$ in the inner heliosheath \citep{ric08,glo11,rei12}.
These ions move through the neutral hydrogen with $n_{H}$ = 0.1 cm$^{-3}$ whose speed is only
 25 km s$^{-1}$ relative to the Sun \citep{mcc15}. Therefore, 
\begin{equation}
l_{neutr} = \frac {1}{n_H \sigma}\approx 220 \textup{ au}.
\label{eq:neutralizationlength}
\end{equation} 



The cooling and neutralization lengths of typically 200 au agree for the North, South, and Port Lobe regions with
the dimension derived from $P\times l$ within the respective error bars (this study) and
derived from time variations of ENAs \citep{rei16}. Both methods
rely on ENA measurements and thus can only sample the cooling length of the plasma rather than the full
heliosheath thickness. However, \citet{rei16} also noted that these cooling lengths derived from ENA observations
are similar to model predictions for the full heliosheath thickness, equal 
to 100 au \citep{pog13} up to 240 au \citep{izm09} for the North pole region. 
This implies that in these directions IBEX may sample plasma all the way from the termination 
shock to the heliopause. The pressure method can be used to derive the total 
heliosheath thickness there if the contributions from low energies around 100 eV are 
taken into account. Otherwise, the total pressure 
is underestimated, as shown by \citet{sch11} 
($P \times l = 72 $ pdyn cm$^{-2}$ au for the IBEX-Hi energy range).
 
The ENA measurements from the Central Tail region suggest that the local neutral hydrogen density
is lower than the hitherto assumed 0.1 cm$^{-3}$: the emission thickness derived from $P \times l$
exceeds the cooling and neutralization lengths by a factor of $1.7\pm0.4$. 
The excess cannot be remedied
by major contributions from energies below 80 eV to the total pressure. This would imply typical energies of 15 or 50 eV 
and thus stretch the cooling length to roughly 400 au. But this assumption would also increase
$P \times l$ (Equation~\ref{eq:correctionfactor}). A region of depleted neutral hydrogen density $n_{H}<0.1$ cm$^{-3}$
extending into the heliosheath around the downwind direction offers the simplest solution to the discrepancy
(refer to Equations \ref{eq:coolinglength} and \ref{eq:neutralizationlength}).
The existence of such a depletion is predicted by state of the art global models of the heliosphere 
(e.g., \citet{izm09,hee14}). The magnitude of the depletion depends on the strength of the 
interstellar magnetic field \citep{hee14}.

The ENA spectra presented here suggest that the neutral hydrogen density in 
Central Tail direction cannot exceed $0.1\times 210$ au / 280 au, and more likely
drops to 0.06 cm$^{-3}$ if we rely on the 8-year average $P\times l= 466$ pdyn cm$^{-2}$ au 
and $P =1.3$ pdyn cm$^{-2}$. The lower limit could be as low as 0.02 cm$^{-3}$ if we apply
the upper limit of roughly 1000 au from Table~\ref{tab:pressure}. This range of possible
neutral densities agrees well with the $\sim 0.06$ cm$^{-3}$ predicted by the simulation of \citet{hee14} for the
central heliotail for an interstellar magnetic field strength of 3 $\mu$G. 
This field strength is consistent with the $2.93\pm0.08$ $\mu$G determined by \citet{zir16b} based on 
the geometry of the ENA Ribbon.

\citet{sch11} argued that since the ENA emission thickness cannot exceed the cooling length, the 
dimension implied by ENA observations was just the cooling length, and $d_{TS}$ in turn had to be smaller 
than 145 au. But with our new analysis this argument does not limit the termination shock distance any
more to $d_{TS}<200$ au. 

Our analysis yields a total plasma pressure outside the termination shock of
1.7 pdyn cm$^{-2}$ with a lower limit of 1.0 pdyn cm$^{-2}$ towards the heliotail. 
Inside the termination shock, the total pressure of the solar wind is dominated by the ram term, with
the internal pressure of pickup ion and magnetic pressure the next two smaller contributions at heliocentric
distances beyond 50 au \citep{wha98}.
Let us just compare the ram pressure term with the lower limit of the heliosheath plasma pressure. 
Because the solar wind density, and thus the ram pressure, drops with the square of the heliocentric distance,
the termination shock must be rather close to the Sun also for the downwind hemisphere.
We obtain the required minimum pressure of 
\begin{equation}
P = \frac{n_{p} m_{H}}{2}v_{sw}^2 = 0.1 \textup{ pP} = 1.0 \textup{ pdyn cm}^{-2}
\end{equation}
with a total proton density of $n_{p} = 10^{-3}$ cm$^{-3}$ and a bulk flow speed of 340 km s$^{-1}$, which
are typical values at 100 au distance \citep{wha98}.
This is consistent with the 115 au derived for 1.0~pdyn~cm$^{-2}$ by \citet{sch11} who also included internal pressure. 
These distances are similar to $d_{TS}$ = 130 au at the north pole and $d_{TS}$ = 110 au at the south pole \citep{rei16}.
In summary, IBEX observations argue for a rather spherical shape of the termination shock (similar to the
model of \citet{izm15}, e.g.).  
 

\section{Conclusions}\label{sec:conclusions}

We have aggregated all observations of heliospheric ENAs from the downwind
hemisphere at low energies measured during the first 8 years of the IBEX mission. 

\begin{itemize}
\item We confirmed previous studies on heliospheric ENAs: Spectral index of ENA intensity 
depends on heliolatitude at solar wind energy, but this heliolatitudinal ordering disappears 
below 0.9 keV. The ENA energy spectrum probably has a knee around 0.8 keV and rolls over around 0.1 keV
in all downwind directions.
\item We have seen the first indication of temporal changes for low energy ENAs (0.1 and 0.2 keV). The
apparent anti-correlation with solar cycle must be re-visited once a full solar cycle has been covered. 
At solar wind energy (0.4 -- 2.5 keV) 
IBEX-Lo data are insufficient to verify the 30\% changes with time observed with IBEX-Hi.
\item The ISN hydrogen signal recovered already in 2014 during solar maximum conditions, earlier than expected.
\item Composite energy spectra from 10 eV to 6 keV for the dynamic pressure times 
ENA emission thickness in the heliosheath have been compiled for the first 8 years of IBEX-Lo and 7 years of IBEX-Hi data. 
In the downwind hemisphere, the protons giving rise to heliospheric ENAs around 0.1 keV
dominate the total plasma pressure. The study of these ENAs is therefore crucial to understand
the vast regions of the heliosphere where no in-situ observations are available. 
\item  Our observations at low energies confirm that the heliosheath towards Southern latitudes is compressed compared
to Northern latitudes and to the Port Lobe.
\item The dynamic pressure of the plasma in the heliosheath 
reaches 1.7 pdyn cm$^{-2}$ integrated from 10 eV to 6 keV
for any direction in the downwind hemisphere; the lower limit is 1.0 pdyn~cm$^{-2}$. 
\end{itemize}

As a consequence the thickness of the plasma structures responsible 
for emission of ENAs in the heliosheath reach 150--210 au towards the poles and the flanks, 
which is similar to the cooling length of the plasma. Since these dimensions
agree with other observation methods and with model predictions for the total heliosheath thickness, 
IBEX possibly samples ENAs from plasma all the way from the termination shock to the heliopause in all directions 
except the Central Tail. Here, the plasma pressure from ENA spectra implies an ENA emission 
thickness of at least 280 au. This region coincides with the nominal downwind direction 
around $\lambda_{ecl}=76^{\circ}$ longitude. The heliosheath therefore is extended
around the downwind direction compared to the flanks (by at least a factor of 1.4). The 
upper limit of this shape factor is ill constrained: ENA intensities measured below 0.1 keV from anti-ram
direction are affected by large uncertainties and the heliosheath
could be thicker than the plasma cooling length in this region. The derived 
ENA emission thickness along the IBEX line-of-sight indicates that the neutral hydrogen 
is depleted towards the heliotail with respect to 
other heliosheath regions to densities between 0.02 and 0.075 cm$^{-3}$.

We are eagerly waiting for the next three years of IBEX observations until 2019, as this will allow us to cover 
the temporal evolution of heliospheric ENAs over one full solar cycle. The present data suggest that
the ENA intensities around 0.1 keV from the downwind hemisphere anti-correlate with solar activity. If confirmed,
this implies periodic changes in the plasma pressure and/or the heliosheath dimensions
in the downwind hemisphere.  
 
\textit{Acknowledgements.}
We thank all of the outstanding men and women who have made the IBEX mission such a wonderful success.
A.G. and P.W. thank the Swiss National Science foundation for financial support.
M.B., M.A.K., and J.M.S. were supported by Polish National Science Center grant 2015-18-M-ST9-00036.
H.K., E.M., N.S., H.O.F., S.A.F., and D.J.M. were supported by the NASA Explorer program as a part of the IBEX mission.

\clearpage
\clearpage

\clearpage

\clearpage

\clearpage

\clearpage

\clearpage

\clearpage

\clearpage

\clearpage

\begin{table}
\begin{center}
\caption{Ubiquitous background hydrogen count rates in the lower energies of IBEX-Lo
for the years 2013--2016 after the post-acceleration of IBEX-Lo was reduced. The background was quantified
by demanding that the heliospheric ENA signal in the solar inertial reference frame
(between 300$^{\circ}$ and 360$^{\circ}$ ecliptic longitude)
should be equal for ram and anti-ram observations after background subtraction.}\label{tab:background_postPACchange}
\begin{tabular}{llr}
\tableline
energy bin & center energy & background count rate in s$^{-1}$\\
\tableline
1 & 0.015 keV & $0.0067\pm0.0015$\\
2 & 0.029 keV & $0.0075\pm0.001$\\
3 & 0.055 keV & $0.0076\pm0.0018$\\
4 & 0.110 keV & $0.0074\pm0.002$\\
5 & 0.209 keV & $0.0012\pm0.0012$\\
6 & 0.439 keV & $0.0002\pm0.0002$\\
7 & 0.872 keV & Not detectable\\
8 & 1.821 keV & Not detectable\\
\tableline
\end{tabular}
\end{center}
\end{table}

\begin{table}
\begin{center}
\caption{The implications of single count limits on corrected ENA intensity.}\label{tab:singlecountlimits}
\begin{tabular}{lllr}
\tableline
energy bin & ecliptic latitude & relative motion of spacecraft & corrected ENA intensity\\
 &  &  &in cm$^{-2}$ sr$^{-1}$ s$^{-1}$ keV$^{-1}$\\
\tableline
1 & 0$^{\circ}$ & ram & 1000 \\
1 & 45$^{\circ}$ & ram & 300 \\
1 & 0$^{\circ}$ & anti-ram & $10^{6}$ \\
1 & 45$^{\circ}$ & anti-ram & 2500 \\
2 & 0$^{\circ}$ & ram & 80 \\
2 & 45$^{\circ}$ & ram & 80 \\
2 & 0$^{\circ}$ & anti-ram & 2000 \\
2 & 45$^{\circ}$ & anti-ram & 250 \\
3 & 0$^{\circ}$ & ram & 56 \\
3 & 45$^{\circ}$ & ram & 45 \\
3 & 0$^{\circ}$ & anti-ram & 200 \\
3 & 45$^{\circ}$ & anti-ram & 110 \\
\tableline
\end{tabular}
\end{center}
\end{table}	

\begin{table}
\begin{center}
\caption{Energy spectrum of ENA intensities in cm$^{-2}$ sr$^{-1}$ s$^{-1}$ keV$^{-1}$ sampled in the four downwind regions, 
averaged over all 8 years of available IBEX-Lo observations.}\label{tab:energyspectra}
\begin{footnotesize}
\begin{tabular}{llllll}
\tableline
Central energy & \textbf{North} & \textbf{South} & \textbf{Central} & \textbf{Port} \\
(keV) &  & & \textbf{Tail} & \textbf{Lobe} \\
\tableline
0.015  & $0 \begin{array}{rr}+29,000 \\-0\end{array}$ & N/A & $0 \begin{array}{rr}+25,000 \\-0\end{array}$ & $0 \begin{array}{rr}+30,000 \\-0\end{array}$ \\
0.029   & $0 \begin{array}{rr}+1700 \\-0\end{array}$ & $3400 \begin{array}{rr}+4000 \\-3400\end{array}$ & $5200 \begin{array}{rr}+84,000 \\-5200\end{array}$ & $0 \begin{array}{rr}+7200 \\-0\end{array}$ \\
0.055   & $2800 \begin{array}{rr}+2000 \\-2200\end{array}$ & $1400 \begin{array}{rr}+600 \\-1300\end{array}$ & $3800 \begin{array}{rr}+15,500 \\-3800\end{array}$ & $2200 \begin{array}{rr}+1600 \\-2200\end{array}$\\
0.110   & $3500 \begin{array}{rr}+300 \\-1500\end{array}$ & $1500 \begin{array}{rr}+500 \\-600\end{array}$ & $6700 \begin{array}{rr}+3400 \\-4000\end{array}$ & $3500 \begin{array}{rr}+1200 \\-700\end{array}$ \\
0.209   & $1620 \begin{array}{rr}+960 \\-410\end{array}$ & $800 \begin{array}{rr}+820 \\-290\end{array}$ & $2200 \begin{array}{rr}+1000 \\-840\end{array}$ & $1450 \begin{array}{rr}+530 \\-300\end{array}$ \\
0.439   & $300 \pm 90$ & $220 \pm 70 $ & $410 \pm 120$ & $270 \pm 80$ \\
0.872   & $138 \pm 41$ & $92 \pm 28 $ & $168 \pm 50$ & $118 \pm 35$ \\
1.821   & $73 \pm 22$ & $47 \pm 14 $ & $63 \pm 19$ & $53 \pm 16$ \\
\tableline
\end{tabular}
\end{footnotesize}
\end{center}
\end{table}	

\begin{table}
\begin{center}
\caption{Trace-back times in years for ENAs measured in IBEX-Lo energy bins for the 5 cases relevant to this study.}\label{tab:trace-back}
\begin{footnotesize}
\begin{tabular}{llllll}
\tableline
Central energy & \textbf{North} & \textbf{South} & \textbf{Port Lobe,} & \textbf{Port Lobe,} & \textbf{Central Tail,} \\
(keV) &  & & \textbf{ram} & \textbf{antiram} & \textbf{antiram}\\
\tableline
0.015  & $20\pm1$ & $16\pm1$ & $14\pm1$ & $36\pm5$ & $43\pm7$\\
0.029   & $15\pm1$ & $12\pm1$ & $12\pm1$ & $23\pm1$ & $27\pm2$\\
0.055   & 12 & 10 & 9.3 & 15 & 18 \\
0.110   & 10 & 8.0 & 7.9 & 11 & 13 \\
0.209   & 7.8 & 6.2 & 6.6 & 8.2 & 9.7 \\
0.439   & 6.0 & 4.8 & 5.4 & 6.1 & 7.2 \\
0.872   & 4.8 & 3.9 & 4.6 & 4.9 & 5.8 \\
1.821   & 4.0 & 3.2 & 3.9 & 4.1 & 4.8 \\
\tableline
\end{tabular}
\end{footnotesize}
\end{center}
\end{table}					

\begin{table}
\begin{center}
\caption{The dynamic pressures times ENA emission thickness $\Delta P\times l$ (Equation\ref{eq:pressurebalance}) 
derived from the composite IBEX-Lo (taken from Table \ref{tab:energyspectra})
 and IBEX-Hi \citep{mcc17} spectra for the four downwind regions
for all available 8 years (7 years in the case of IBEX-Hi). 
For the correction factor $c_f$ in the second column (Eq.~\ref{eq:correctionfactor}) a constant 140 km s$^{-1}$ was assumed everywhere.}\label{tab:pressurespectrum}
\begin{footnotesize}
\begin{tabular}{llllll}
\tableline
Central energy & Correction & \textbf{North} & \textbf{South} & \textbf{Central} & \textbf{Port} \\
(keV) & factor &  & & \textbf{Tail} & \textbf{Lobe} \\
  &  & $\Delta P\times l$ & (pdyn  & cm$^{-2}$ & au) \\
\tableline
0.015 & 170 & $0 \begin{array}{rr}+317 \\-0\end{array}$ & N/A & $0 \begin{array}{rr}+267 \\-0\end{array}$ & $0 \begin{array}{rr}+327 \\-0\end{array}$ \\
0.029 & 69  & $0 \begin{array}{rr}+23 \\-0\end{array}$ & $49 \begin{array}{rr}+54 \\-45\end{array}$ & $71 \begin{array}{rr}+1124 \\-71\end{array}$ & $0 \begin{array}{rr}+97 \\-0\end{array}$ \\
0.055 & 31  & $50 \begin{array}{rr}+36 \\-40\end{array}$ & $25 \begin{array}{rr}+36 \\-24\end{array}$ & $68 \begin{array}{rr}+281 \\-68\end{array}$ & $39 \begin{array}{rr}+29 \\-39\end{array}$\\
0.110 & 15  & $102 \begin{array}{rr}+9 \\-41\end{array}$ & $42 \begin{array}{rr}+16 \\-13\end{array}$ & $205 \begin{array}{rr}+95 \\-110\end{array}$ & $103 \begin{array}{rr}+34 \\-20\end{array}$ \\
0.209 & 8.3 & $87 \begin{array}{rr}+44 \\-19\end{array}$ & $41 \begin{array}{rr}+37 \\-13\end{array}$ & $118 \begin{array}{rr}+45 \\-38\end{array}$ & $77 \begin{array}{rr}+24 \\-14\end{array}$ \\
0.439 & 4.8 & $31 \begin{array}{rr}+9 \\-9\end{array}$ & $22 \begin{array}{rr}+6 \\-6\end{array}$ & $42 \begin{array}{rr}+12 \\-12\end{array}$ & $27 \begin{array}{rr}+8 \\-8\end{array}$ \\
0.71 & 3.6  & $22 \begin{array}{rr}+4 \\-4\end{array}$ & $15 \begin{array}{rr}+3 \\-3\end{array}$ & $26 \begin{array}{rr}+10 \\-5\end{array}$& $23 \begin{array}{rr}+5 \\-5\end{array}$\\
1.11 & 2.9  & $23 \begin{array}{rr}+5 \\-5\end{array}$ & $15 \begin{array}{rr}+3 \\-3\end{array}$ & $29 \begin{array}{rr}+6 \\-8\end{array}$ & $20 \begin{array}{rr}+8 \\-4\end{array}$ \\
1.74 & 2.4  & $22 \begin{array}{rr}+8 \\-5\end{array}$ & $13 \begin{array}{rr}+3 \\-3\end{array}$ & $23 \begin{array}{rr}+5 \\-8\end{array}$& $14 \begin{array}{rr}+8 \\-3\end{array}$ \\
2.73 & 2.0  & $22 \begin{array}{rr}+4 \\-6\end{array}$ & $14 \begin{array}{rr}+3 \\-3\end{array}$ & $14 \begin{array}{rr}+3 \\-4\end{array}$& $8 \begin{array}{rr}+5 \\-2\end{array}$ \\
4.29 & 1.8  & $16 \begin{array}{rr}+3 \\-3\end{array}$ & $13 \begin{array}{rr}+3 \\-3\end{array}$ & $9 \begin{array}{rr}+3 \\-3\end{array}$ & $6 \begin{array}{rr}+3 \\-2\end{array}$\\
\tableline
\end{tabular}
\end{footnotesize}
\end{center}
\end{table}

\begin{table}
\begin{center}
\caption{Sums of the dynamic pressures times ENA emission thickness from 
Table~\ref{tab:pressurespectrum} (first two lines) and the lower limits for high solar activity conditions 
and best signal-to-noise ratio (bottom line, bold).}\label{tab:pressure}
\begin{footnotesize}
\begin{tabular}{llllll}
\tableline
Energy range & Observation & \textbf{North} & \textbf{South} & \textbf{Central} & \textbf{Port} \\
(keV) & time &  & & \textbf{Tail} & \textbf{Lobe} \\
  &  & $P \times l$ & (pdyn  & cm$^{-2}$ & au) \\
\tableline
0.01 -- 6.0 & 2009--2016 & $374 \begin{array}{rr}+323 \\-62\end{array}$  & $248 \begin{array}{rr}+77 \\-55\end{array}$ & $605 \begin{array}{rr}+1194 \\-153\end{array}$ & $318 \begin{array}{rr}+345 \\-47\end{array}$\\
0.08 -- 6.0 & 2009--2016 & $325 \begin{array}{rr}+47 \\-47\end{array}$ & $174 \begin{array}{rr}+42 \\-21\end{array}$ & $466 \begin{array}{rr}+107 \\-118\end{array}$ & $279 \begin{array}{rr}+45 \\-26\end{array}$ \\
\tableline
0.01 -- 6.0 & 2009--2011 & \textbf{210} & \textbf{150} & \textbf{280} & \textbf{180} \\
\tableline
\end{tabular}
\end{footnotesize}
\end{center}
\end{table}


\begin{thebibliography}{}



\bibitem[Burlaga et al.(2008)]{bur08} Burlaga, L. F., Ness, N. F., Acu\~{n}a, M. H., et al.
2008, Nature, 454, 75

\bibitem[Bzowski(2008)]{bzo08} Bzowski, M. 2008, Astronomy \& Astrophysics, 488, 1057





\bibitem[Bzowski et al.(2017)]{bzo17} Bzowski, M., Kubiak, M. A., Czechowski, A., \& Grygorczuk, J. 2017, The Astrophysical Journal, 845, 15



\bibitem[Dialynas et al.(2017)]{dia17} Dialynas, K., Krimigis, S. M., Mitchell, D. G., Decker, R. B., \& Roelof, E. C., 2017,
Nature Astronomy, 1, 115

\bibitem[Decker et al.(2005)]{dec05} Decker, R. B., Krimigis, S. M., Roelof, E. C., et al. 2005, Science, 309, 2020


\bibitem[Desai et al.(2015)]{des15} Desai, M. I., Allegrini, F., Dayeh, M. A., et al. 2015, The Astrophysical Journal, 802, 100



\bibitem[Funsten et al.(2009)]{fun09} Funsten, H. O., Allegrini, F., Bochsler, P., 
et al. 2009, Space Science Reviews, 146, 75

\bibitem[Fuselier et al.(2009)]{fus09} Fuselier, S. A., Bochsler, P., Chornay, D.,
et al. 2009, Space Science Reviews, 146, 117


\bibitem[Fuselier et al.(2012)]{fus12} Fuselier, S. A., Allegrini, F.,
Bzowski, M., et al. 2012, The Astrophysical Journal, 754, 14

\bibitem[Fuselier et al.(2014)]{fus14} Fuselier, S. A., Allegrini, F.,
Bzowski, M., et al. 2014, The Astrophysical Journal, 784, 89

\bibitem[Galli et al.(2013)]{gal13} Galli, A., Wurz, P., Kollmann, P., et al.
2013, The Astrophysical Journal, 775, 24

\bibitem[Galli et al.(2014)]{gal14} Galli, A., Wurz, P., Fuselier, S. A., et al.
2014, The Astrophysical Journal, 796, 9

\bibitem[Galli et al.(2015)]{gal15} Galli, A., Wurz, P., Park, J., et al.
2015, The Astrophysical Journal, 220, 30

\bibitem[Galli et al.(2016)]{gal16} Galli, A., Wurz, P., Schwadron, N.A., et al.
2016, The Astrophysical Journal, 821, 107

\bibitem[Gloeckler \& Fisk(2011)]{glo11} Gloeckler, G. \& Fisk, L. A. 2011, in AIP Conf. Proc. 1302, Pickup Ions
throughout the Heliosphere and Beyond, ed. J. A. Le Roux, V. Florinski, G.
P. Zank, \& A. J. Coates (Melville, NY: AIP), 110, doi: http://dx.doi.org/10.1063/1.3529957

\bibitem[Gloeckler \& Fisk(2015)]{glo15} Gloeckler, G. \& Fisk, L. A. 2015, The Astrophysical Journal Letters, 806, L27

\bibitem[Gurnett et al.(2013)]{gur13} Gurnett, D. A., Kurth, W. S., Burlaga, L. F., \& Ness, N. F. 2013,
Science, 341, 1489


\bibitem[Heerikhuisen et al.(2014)]{hee14} Heerikhuisen, J., Zirnstein, E. J., Funsten, H. O., Pogorelov, N. V., \& Zank, G.
P. 2014, The Astrophysical Journal, 784, 73


\bibitem[Izmodenov et al.(2009)]{izm09} Izmodenov, V. V., Malama, Y. G., Ruderman, M. S., et al. 2009, 
Space Science Reviews, 146, 329

\bibitem[Izmodenov and Alexashov(2015)]{izm15} Izmodenov, V. V. \& Alexashov, D. B., 2015, 
The Astrophysical Journal Supplement Series, 220, 32



\bibitem[Kubiak et al.(2014)]{kub14} Kubiak, M. A., Bzowski, M., Sok\'{o}{\l}, J. M., et al. 2014, The Astrophysical Journal Supplement Series,
213, 29

\bibitem[Kubiak et al.(2016)]{kub16} Kubiak, M. A., Swaczyna, P., Bzowski, M., et al. 2016, The Astrophysical Journal Supplement Series, 223, 25


\bibitem[Lindsay \& Stebbings(2005)]{lin05} Lindsay, B. G. \& Stebbings, R. F. 2005, Journal of Geophysical Research, 110, A12213

\bibitem[Livadiotis et al.(2013)]{liv13} Livadiotis, G., McComas, D. J., Schwadron, N. A., Funsten, H. O., \& Fuselier,
S. A. 2013, The Astrophysical Journal, 762, 134


\bibitem[McComas et al.(2009)]{mcc09} McComas, D.J., Allegrini, F., Bochsler, P.,
et al. 2009, Space Science Reviews, 146, 11




\bibitem[McComas et al.(2013)]{mcc13} McComas, D. J., Dayeh, M. A., Funsten, H. O., 
Livadiotis, G., \& Schwadron, N. A. 2013, The Astrophysical Journal, 771, 77


\bibitem[McComas et al.(2014)]{mcc14} McComas, D. J., Lewis, W. S., \& Schwadron, N. A. 2014,
Reviews of Geophysics, 52, doi:10.1002/2013RG000438

\bibitem[McComas et al.(2015)]{mcc15} McComas, D. J., Bzowski, M., Fuselier, S. A., et al. 2015, The Astrophysical Journal Supplement Series, 220, 22

\bibitem[McComas et al.(2017)]{mcc17} McComas, D. J., Zirnstein, E. J., Bzowski, M., et al. 2017, The Astrophysical Journal Supplement Series, 229, 41

\bibitem[M\"{o}bius et al.(2012)]{moe12} M\"{o}bius, E., Bochsler, P., Bzowski, M. et al. 2012,
The Astrophysical Journal Supplement Series, 198, 11


\bibitem[Park et al.(2016)]{par16} Park, J., Kucharek, H., M\"{o}bius, E. et al. 2016, The Astrophysical Journal, 833, 130

\bibitem[Pogorelov et al.(2013)]{pog13} Pogorelov, N. V., Seuss, S. T., Borovikov, S. N., et al. 2013, The Astrophysical Journal, 772, 2

\bibitem[Reisenfeld et al.(2012)]{rei12} Reisenfeld, D. B., Allegrini, F., Bzowski, M. et al.
2012, The Astrophysical Journal, 747, 110

\bibitem[Reisenfeld et al.(2016)]{rei16} Reisenfeld, D. B., Bzowski, M., Funsten, H.O. et al.
2016, The Astrophysical Journal, 833, 277

\bibitem[Richardson et al.(2008)]{ric08} Richardson, J. D., Kasper, J. C., Wang, et al.
2008, Nature, 454, 63



\bibitem[Saul et al.(2013)]{sau13} Saul, L., Bzowski, M., Fuselier, S. A., et al. 2013, The Astrophysical Journal, 767, 130 

\bibitem[Schwadron et al.(2011)]{sch11} Schwadron, N. A., Allegrini, F., Bzowski, M., et al.
2011, The Astrophysical Journal, 731, 56


\bibitem[Schwadron et al.(2014)]{sch14} Schwadron, N. A., M\"{o}bius, E., Fuselier, S. A. et al.
2014, The Astrophysical Journal, 215, 13




\bibitem[Sok\'{o}{\l} et al.(2015)]{sok15} Sok\'{o}{\l}, J. M., Kubiak, M. A., Bzowski, M., Swaczyna, P., 
et al. 2015a, The Astrophysical Journal Supplement Series, 220, 27

\bibitem[SILSO, World Data Center(2008)]{wdc08} SILSO, World Data Center
-- Sunspot Number and Long-term Solar Observations,
Royal Observatory of Belgium 2008--2014,
International Sunspot Number Monthly Bulletin and online catalogue, http://www.sidc.be/SILSO/

\bibitem[Stone et al.(2013)]{sto13} Stone, E. C., Cummings, A. C., McDonald, F. B., et al. 2013, Science, 341, 15


\bibitem[Whang et al.(1998)]{wha98} Whang Y. C. 1998, Journal of Geophysical Research, 103, 17419

\bibitem[Whang et al.(1999)]{wha99} Whang Y. C., Lu, J. Y., Burlaga L. F. 1999, Journal of Geophysical Research, 104, 28255



\bibitem[Wood et al.(2014)]{woo14} Wood, B. E., Izmodenov, V. V., Alexashov, D. B., Redfield, S., \& Edelman, E. 2014, 
The Astrophysical Journal, 780, 108




\bibitem[Zirnstein et al.(2016)]{zir16} Zirnstein, E. J., Funsten, H. O., Heerikhuisen, J., McComas, D. J., 
Schwadron, N. A., \& Zank, G. P. 2016, The Astrophysical Journal, 826, 58

\bibitem[Zirnstein et al.(2016b)]{zir16b} Zirnstein, E. J., Heerikhuisen, J., Funsten, H. O., 
Livadiotis, G., McComas, D. J., \& Pogorelov, N. V. 2016b, The Astrophysical Journal Letters, 818, L18 

\bibitem[Zirnstein et al.(2017)]{zir17} Zirnstein, E. J., Dayeh, M. A., McComas, D. J., \& Sok\'{o}{\l}, J.M. 2017, 
The Astrophysical Journal, 846, 63

\end{thebibliography}
\end{document}